\documentclass[prx,twocolumn,english,superscriptaddress,floatfix,longbibliography]{revtex4-2}

\usepackage{amsthm}
\usepackage{graphicx}% Include figure files
\usepackage{dcolumn}% Align table columns on decimal point
\usepackage{bm}% bold math
\usepackage[utf8]{inputenc}
\usepackage{mathptmx}
\usepackage{multirow}
\usepackage{braket}
\usepackage[english]{babel}
\usepackage{xcolor}% Include colors for document elements
\colorlet{RED}{red}
\colorlet{black}{black}
\usepackage{dcolumn}% Align table columns on decimal pointhttps://v1.overleaf.com/17481932tmwjwzytcbcq#
\usepackage{bm}% bold math
\usepackage[version=4]{mhchem} % HvD: for general chemical formula like \ce{(TiO2)n}
\usepackage{acronym}
\usepackage{adjustbox}
\usepackage{float}
\usepackage{microtype} % fixes many of the overflow hbox errors
\definecolor{background-color}{gray}{0.98}
\usepackage[margin=2.3cm,bmargin=1cm,footnotesep=1cm]{geometry}
\usepackage{pgfplots}
\usetikzlibrary{intersections}
\usepgfplotslibrary{fillbetween}
\pgfplotsset{compat=1.18}
\usepackage{physics}
\usepackage{bm}% bold math
\usepackage{xcolor}
\usepackage{wrapfig}
\usepackage{float}
\usepackage{tikz}
\usetikzlibrary{arrows.meta, positioning, fit}

\usepackage{enumitem}
\usepackage{makecell}

\usepackage[most]{tcolorbox}
\tcbuselibrary{skins,breakable,hooks,theorems}

\newtcolorbox[auto counter]{DeepDive}[3][]{%
  enhanced, breakable, colback=black!2, colframe=black, boxrule=0.4pt, arc=1mm,
  left=1.2ex, right=1.2ex, top=0.8ex, bottom=0.8ex,
  fonttitle=\bfseries,
  title=Deep--Dive~{#3}: #2, % title arg
  label={#3},
}
\newlist{proscons}{itemize}{1}
\setlist[proscons]{label={}, leftmargin=1.2em, nosep}

\newcommand{\UWChem}{Department of Chemistry, Yale University, New Haven, Connecticut 06520, United States}
\newcommand{\PCSD}{Physical and Computational Sciences Directorate, Pacific Northwest National Laboratory, Richland, Washington 99354, United States}
\newcommand{\ORNL}{Quantum Information Science Section, Oak Ridge National Laboratory, Oak Ridge, Tennessee 37831, United States}
\newcommand{\YaleChem}{Department of Chemistry, Yale University, New Haven, Connecticut 06520, United States}
\newcommand{\UChicago}{Department of Chemistry, Chicago Center for Theoretical Chemistry, University of Chicago, Chicago, Iillinois 60637, United States}
\newcommand{\PNNLCloud}{Center for Cloud Computing, Pacific Northwest National Laboratory, Richland, Washington 99354, United States}
\newcommand{\UWPhys}{Department of Physics, University of Washington, Seattle, Washington 98195, USA}
\newcommand{\UWCS}{Department of Electrical and Computer Engineering, University of Washington, Seattle, Washington 98195, United States}

\begin{document}

\title{A Perspective on Quantum Computing Applications in Quantum Chemistry using 25--100 Logical Qubits}

\author{Yuri Alexeev}
\affiliation{NVIDIA Corporation, Santa Clara, California 95051, United States}

\author{Victor S. Batista}
\affiliation{\YaleChem}
\affiliation{Yale Quantum Institute, Yale University, New Haven, Connecticut 06511, United States}

\author{Nicholas Bauman}
\affiliation{\PCSD}

\author{Luke Bertels}
\affiliation{\ORNL}

\author{Daniel Claudino}
\affiliation{\ORNL}

\author{Rishab Dutta}
\affiliation{\YaleChem}
\affiliation{\PCSD}

\author{Laura Gagliardi}
\affiliation{\UChicago}

\author{Scott Godwin}
\affiliation{\PNNLCloud}

\author{Niranjan Govind}
\affiliation{\PCSD}
\affiliation{\UWChem}

\author{Martin Head-Gordon}
\affiliation{Department of Chemistry, University of California, Berkeley, California 94720, United States}

\author{Matthew Hermes}
\affiliation{\UChicago}

\author{Karol Kowalski}
\affiliation{\PCSD}
\affiliation{\UWPhys}

\author{Ang Li}
\affiliation{\PCSD}
\affiliation{\UWCS}

\author{Chenxu Liu}
\affiliation{\PCSD}

\author{Junyu Liu}
\affiliation{Department of Computer Science, School of Computing \& Information, University of Pittsburgh, Pittsburgh, Pennsylvania 15260, United States}

\author{Ping Liu}
\affiliation{Chemistry Division, Brookhaven National Laboratory, Upton, New York 11973, United States}

\author{Juan M. Garc\'{i}a-Lustra}
\affiliation{Department of Energy Conversion and Storage, Technical University of Denmark, DK-2800 Kgs Lyngby, Denmark}

\author{Daniel Mejia-Rodriguez}
\affiliation{\PCSD}

\author{Karl Mueller}
\affiliation{\PCSD}

\author{Matthew Otten}
\affiliation{Department of Physics, University of Wisconsin -- Madison, Madison, Wisconsin 53706, United States}

\author{Bo Peng}
\email{The authors are listed in alphabetical order by last name. All authors contributed equally to this work. Correspondence should be addressed to peng398@pnnl.gov.}
\affiliation{\PCSD}

\author{Mark Raugus}
\affiliation{National Security Directorate, Pacific Northwest National Laboratory, Richland, Washington 99354, United States}

\author{Markus Reiher}
\affiliation{Department of Chemistry and Applied Biosciences, ETH Zurich, Vladimir-Prelog-Weg 2, Zurich 8093, Switzerland}

\author{Paul Rigor}
\affiliation{Center for Continuum Computing, Pacific Northwest National Laboratory, Richland, Washington 99354, United States}

\author{Wendy Shaw}
\affiliation{\PCSD}

\author{Mark van Schilfgaarde}
\affiliation{Materials Science Center, National Renewable Energy Laboratory, Golden, Colorado 80401, United States}

\author{Tejs Vegge}
\affiliation{Pioneer Center for Accelerating P2X Materials Discovery, Department of Energy Conversion and Storage, Technical University of Denmark, DK-2800 Kgs Lyngby, Denmark}

\author{Yu Zhang}
\affiliation{Theoretical Division, Los Alamos National Laboratory, Los Alamos, New Mexico 87545, United States}

\author{Muqing Zheng}
\affiliation{\PCSD}

\author{Linghua Zhu}
\affiliation{\UWChem}

\today

\begin{abstract}
The intersection of quantum computing and quantum chemistry represents a promising frontier for achieving quantum utility in domains of both scientific and societal relevance. Owing to the exponential growth of classical resource requirements for simulating quantum systems, quantum chemistry has long been recognized as a natural candidate for quantum computation. This perspective focuses on identifying scientifically meaningful use cases where early fault-tolerant quantum computers, which are considered to be equipped with approximately 25--100 logical qubits, could deliver tangible impact. \textcolor{black}{While recent advances in classical computing have pushed the boundaries of tractable simulations to unprecedented scales, this logical-qubit regime represents the first window where quantum devices can pursue qualitatively distinct strategies, such as polynomial-scaling phase estimation, direct simulation of quantum dynamics, and active-space embedding, that remain challenging for classical solvers, for instance, multireference charge-transfer and conical-intersection states central to photochemistry and materials design.} We highlight near-term opportunities in algorithm and software design, discuss representative chemical problems suited for quantum acceleration, and propose strategic roadmaps and collaborative pathways for advancing practical quantum utility in quantum chemistry.
\end{abstract}

\maketitle

\tableofcontents
%%%%%%%%%%%%% Sec.~I
\section{Introduction and Motivation}

\textcolor{black}{The year of 2025} marks a significant milestone---roughly a century since the formulation of quantum mechanics fundamentally reshaped our understanding of matter at the atomic and molecular level~\cite{Schrodinger1926,Heisenberg1927,Dirac1928}. The Schr\"{o}dinger equation, introduced during this transformative period, provided the theoretical bedrock for quantum chemistry, enabling, in principle, the prediction of chemical properties and reactivity from first principles, famously demonstrated for the hydrogen molecule~\cite{HeitlerLondon1927}. Over the subsequent 100 years, the field has witnessed impressive advances, with the development and application of sophisticated classical computational methods, such as Density Functional Theory (DFT)~\cite{HohenbergKohn1964,KohnSham1965,ParrYang1989} and wave function-based approaches like Coupled Cluster (CC) theory~\cite{Cizek1966,PALDUS2005115,BartlettMusial2007}, achieving remarkable success in explaining and predicting chemical phenomena~\cite{Schaefer2012quantum,Kohn1999nobel}. 

Despite a century of progress and the power of established classical algorithms, substantial challenges remain. The inherent complexity of the quantum many-body problem, the very challenge laid barely by quantum mechanics itself, continues to limit the accuracy and applicability of classical methods, 
\textcolor{black}{particularly in the treatment of:  
(i) strongly correlated electronic systems, such as catalytic sites like FeMoco, where single-reference methods fail~\cite{Reiher2017};  
(ii) complex excited states crucial for photochemistry and materials science, such as conical intersections and charge-transfer states that underlie ultrafast photophysical processes~\cite{DreuwHeadGordon2005};  
(iii) open quantum dynamics that govern system–environment interactions~\cite{BreuerPetruccione2002}; and  
(iv) transition-state energetics and reaction barriers, where stretched bonds induce strong/static correlation beyond single-reference methods, and dispersion or environmental effects can secondarily modulate barrier heights~\cite{Klippenstein2014chemical}. }
These enduring limitations, stemming directly from the exponential growth of the Hilbert space with system size, motivate the exploration of fundamentally different computational paradigms. Quantum computational approaches, which leverage quantum phenomena directly, offer a potential pathway to supplement or even surpass classical techniques by tackling these intrinsically quantum problems more naturally, as envisioned early on by Feynman~\cite{Feynman1982} and Lloyd~\cite{Lloyd1996}, and extensively reviewed ever since~\cite{AspuruGuzik2005,McArdle2020,Bauer2020}.

\textcolor{black}{Recent advances in classical computing have impressively extended the reach of exact and approximate electronic structure methods, including trillion-determinant full configuration interaction~\cite{doi:10.1021/acs.jctc.3c01190} and massively parallel relativistic density matrix renormalization group (DMRG) tailored coupled cluster approach~\cite{doi:10.1021/acs.jctc.4c00641}. These studies approach orbital ranges comparable to those accessible with 25--100 logical qubits under conventional mappings. Yet classical solvers remain subject to exponential barriers for general strongly correlated and dynamical problems. By contrast, even early fault-tolerant devices with 25--100 logical qubits can access qualitatively different algorithmic primitives, such as polynomial-scaling phase estimation, efficient Hamiltonian simulation, and block-encoded dynamics, that cannot be emulated efficiently at scale by classical algorithms.}

\textcolor{black}{Here, following recent usage in the community, we use the term \emph{quantum utility} to mean reliable, validated quantum computations on domain-relevant tasks at scales beyond brute-force classical methods~\cite{IBMQuantumUtility,KimNature2023,HerrmannUtility2023,HoeflerAdvantage2023}. This notion emphasizes practical, auditable performance on scientifically meaningful problems, reported with stated error bars and resource annotations (a standardized record of quantum resource requirements, including logical-qubit counts, circuit depth/$T$-counts, Hamiltonian term norms, and shot budgets). It contrasts with \emph{quantum advantage}, which often refers to worst-case or contrived separations, and with \emph{quantum dominance/supremacy}, which targets formal separations on specially constructed tasks.}

\textcolor{black}{This perspective is motivated by the opportunity to translate these distinctive quantum capabilities into practical chemistry workflows. Achieving utility in the 25--100 logical-qubit regime will depend on resource-aware algorithm design~\cite{Lee2023}, robust hybrid quantum-classical workflows~\cite{Cerezo2021}, and interdisciplinary co-design across algorithms, chemistry, and hardware~\cite{McCaskey2019}. Realizing this opportunity also requires a clear-eyed view of both the burgeoning capabilities of emerging hardware~\cite{IBMRoadmap2022,GoogleLogicalProcessor2024,Bluvstein2024Logical,Aghaee2025Interferometric} and the practical constraints of near-term fault-tolerant platforms. Before turning to specific quantum chemistry applications and algorithms, we briefly overview this hardware landscape (Sec.~\ref{sec:landscape}). Throughout, the 25--100 logical-qubit regime serves as a unifying lens for evaluating algorithmic strategies, benchmarking protocols, and collaborative pathways, with emphasis on resource-aware and auditable progress. The manuscript then proceeds from opportunities in quantum chemical simulations (Sec.~\ref{sec:oppo}), to algorithmic innovations most relevant for this regime (Sec.~\ref{sec:algo}), to hybrid integration pathways (Sec.~\ref{sec:pathways}), and finally to the conclusion and outlook on co-design and collaboration models (Sec.~\ref{sec:conclusion}).}

\begin{figure*}
 \centering
    \includegraphics[width=\linewidth]{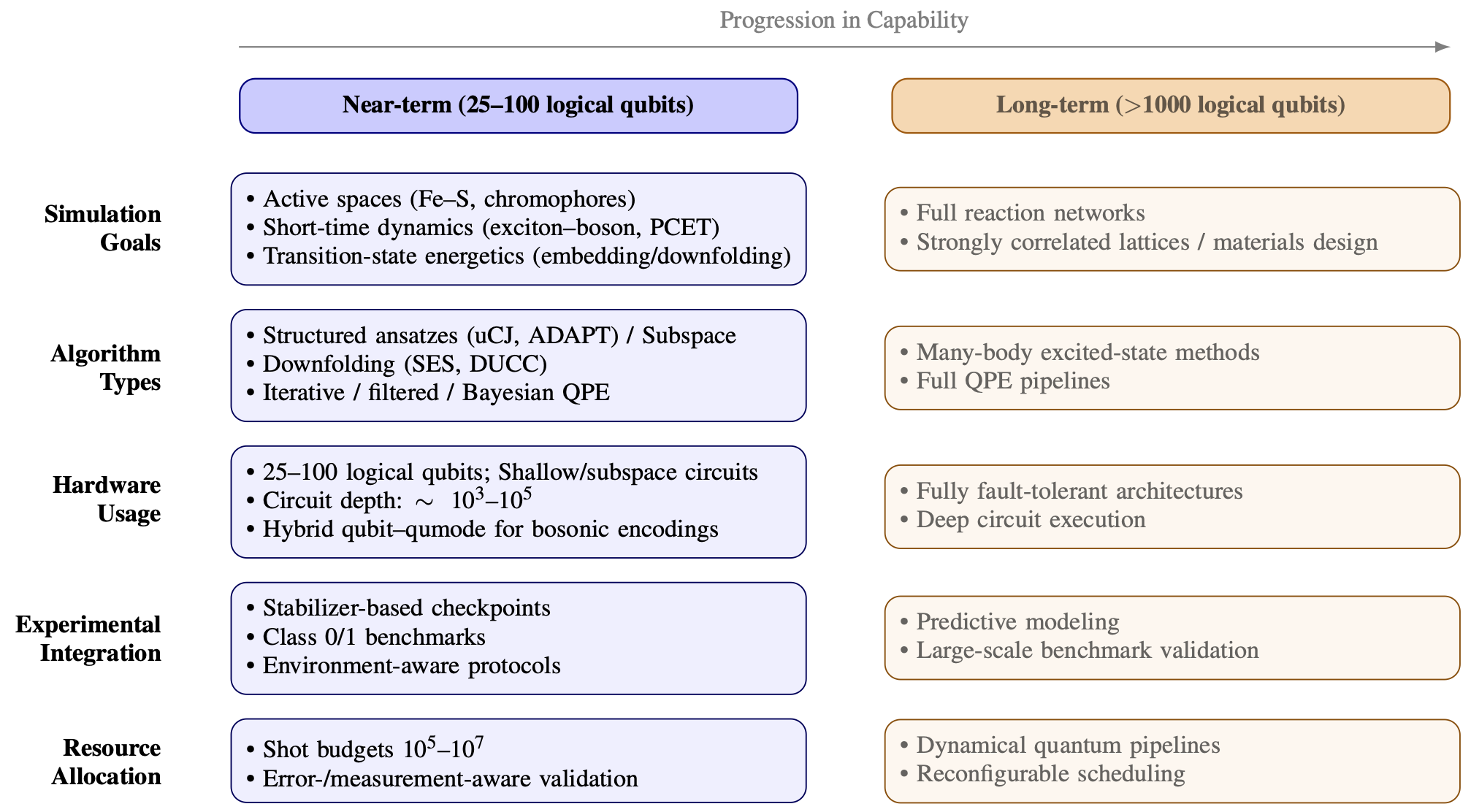}
\caption{\textcolor{black}{Strategic directions for quantum chemical simulations organized by near-term (25--100 logical qubits) and long-term ($>$1000 logical qubits) scales. The near-term column reflects the scope of this Perspective and is grounded in the resource estimates of Table~\ref{tab:resources}. The long-term column is included for context and is not the focus here. Strongly correlated systems are treated as near-term opportunities when approached via compact active spaces (Sec.~\ref{sec:oppo}).}}
\label{fig:strategy}
\end{figure*}

%%%%%%%%%%%%% Sec.~II

\section{The 25--100 logical-qubit regime: A Transitional Landscape}\label{sec:landscape}

\textcolor{black}{\textit{In this section, we outline the evolving quantum hardware landscape and explain why the 25--100 logical-qubit regime marks a pivotal transitional window for quantum chemistry applications.}}
 
Recent advances in quantum hardware have significantly improved the prospects for achieving early fault-tolerant quantum computations. While current NISQ devices have achieved important milestones, including small-scale quantum simulations~\cite{Kandala2017Nature,GoogleSycamoreChem2020} and sophisticated error mitigation~\cite{CaiErrorMitigationReview2023}, scalable quantum utility requires logical qubits protected by quantum error correction (QEC)~\cite{NielsenChuangBook,Preskill2018}.

Implementing a single logical qubit today demands many physical qubits, sometimes thousands, depending on error rates and the chosen code, such as the surface code~\cite{Fowler2012,GidneyEkeraFactoring2021,GoogleSurfaceCode2023}. Nevertheless, improvements in coherence times, gate fidelities, and system integration across leading platforms, including superconducting qubits~\cite{Krinner2022,GoogleLogicalProcessor2024}, atom array~\cite{Bluvstein2024Logical}, trapped ions~\cite{BruzewiczIonTrapReview2019,Pogorelov2021}, and photonic architectures~\cite{MoodyPhotonicsReview2022,PhysRevLett.134.090601}, 
\textcolor{black}{suggest that processors comprising 25–100 logical qubits could plausibly become available on a 5--10 year horizon (see also a recent analysis of use cases from the NERSC workload, Ref.~\citenum{osti_2588210}). This estimate reflects the scaling projections reported in recent national roadmaps~\cite{NASQuantumReport2019,NAP26850}, IBM’s published timeline for error-corrected qubits~\cite{IBMRoadmap2022}, Google’s logical-qubit demonstrations~\cite{GoogleSurfaceCode2023,GoogleLogicalProcessor2024}, and atom-array advances~\cite{Bluvstein2024Logical}. We stress that this horizon is a projection, not a guarantee, and depends critically on continued improvements in coherence, gate fidelity, and decoding throughput.}

In comparison with long-term strategic directions, the 25--100 logical-qubit regime marks a pivotal near-term threshold in the evolution of quantum computing applications in quantum chemistry (see Figure~\ref{fig:strategy}). Devices in this range may be the first to enable meaningful quantum chemistry simulations that are intractable on classical computers, such as accurately modeling complex molecules~\cite{Reiher2017,Lee2023}. 
\textcolor{black}{Figure~\ref{fig:strategy} situates these opportunities within a broader landscape, distinguishing near-term targets from longer-term aspirations. We emphasize that the long-term entries are illustrative and not the focus of this Perspective; our analysis centers on the 25--100 logical-qubit regime, corresponding to the near-term columns. In particular, strongly correlated systems (discussed in detail in Sec.~\ref{sec:oppo}) are highlighted as compelling near-term candidates when treated within compact active spaces, rather than being deferred exclusively to the long-term horizon.}
However, the opportunity comes with substantial caveats: these early fault-tolerant systems will still face non-negligible logical error rates, limited coherence times relative to computation depth, and practical constraints on connectivity, measurement, and classical I/O~\cite{RoettelerSvore2018}.

Progress in this regime demands a re-evaluation of what constitutes ``quantum utility'' in chemistry~\cite{HoeflerAdvantage2023}. It is not solely about outperforming classical solvers in compute time or accuracy for all problems, but rather about delivering new scientific insights into problems that are intrinsically quantum and difficult to treat or beyond classical methods, including strongly correlated electrons~\cite{BartlettMusial2007}, quantum coherence in dynamics~\cite{ChildsDynamicsReview2018}, and environmental interactions~\cite{BreuerPetruccione2002}.
\textcolor{black}{By scientific insights, we mean not just reproducing known quantities more efficiently, but accessing observables and regimes that classical simulations cannot capture reliably. Examples include uncovering mechanistic details of catalytic cycles that hinge on multireference transition states, resolving ultrafast photochemical processes via conical intersections, or quantifying coherence and dissipation effects in biomolecular dynamics. In these cases, even modest quantum devices may provide qualitative understanding, such as relative ordering of states, trends along reaction coordinates, or signatures of coherence, that extend beyond the reach of brute-force classical methods.}
{black}
Development efforts are therefore increasingly centered around hybrid algorithms~\cite{Cerezo2021}, embedding techniques~\cite{SunDMETReview2016,RevModPhys.78.865,negre2025newperspectivesdensitymatrixembedding,Ma2021Embedding,Manby2021Embedding,Jones2020Embedding,Libisch2014Embedded,Jacob2024Subsystem,Wesolowski2015Embedding,PhysRevA.109.022418,waldrop2021projector,waldrop2025projector}, and variational methods~\cite{Peruzzo2014} that operate with shallow circuits or make optimal use of limited logical qubit counts. These strategies aim to optimize the use of limited quantum resources while interfacing seamlessly with classical simulation frameworks.

This regime is particularly well-suited for active space quantum simulations, where a judiciously chosen set of orbitals (often those associated with strong correlation or reactive behavior) is treated on a quantum computer, while classical components handle the weakly correlated environment. Techniques such as downfolding~\cite{KowalskiSubalgebras2018,KowalskiSubsystemFlow2021,KowalskiSubsystemSelfConsistency2023,BaumanVQE_Downfolded2021,PRXQuantum_4_020313,BaumanExcitedStates2019,KowalskiBaumanDynamics2020,BaumanDownfolding2019,BaumanKowalskiJCP2022,BaumanKowalskiMaterialsTheory2022,Metcalf2020} and quantum embedding~\cite{SunDMETReview2016,RevModPhys.78.865,negre2025newperspectivesdensitymatrixembedding,Ma2021Embedding,Manby2021Embedding,Jones2020Embedding,Libisch2014Embedded,Jacob2024Subsystem,Wesolowski2015Embedding,PhysRevA.109.022418,waldrop2021projector,waldrop2025projector} provide viable pathways to construct effective Hamiltonians for such active spaces.

In addition to ground-state energy estimation, quantum dynamics is emerging as an area where early utilities may arise~\cite{Bauer2020,ChildsDynamicsReview2018}. Simulating time-dependent processes, especially in open systems or photoinduced transformations, poses considerable challenges for classical methods due to memory bottlenecks and entanglement growth~\cite{szabo1996modern,helgaker2013molecular}. 
\textcolor{black}{Quantum processors, based on formal complexity bounds for Hamiltonian simulation~\cite{LowHamiltonianSim2017,Berry2019qubitizationof,Low2019hamiltonian}, are projected to enable polynomial-scaling routes for handling such inherently quantum phenomena. Resource estimates in Table~\ref{tab:resources} show that even modest active spaces, such as Fe$_2$S$_2$ clusters or photochemical chromophores, map naturally to 25–40 logical qubits with depth $10^4$–$10^5$ and shot budgets $10^6$–$10^7$, providing concrete targets for this regime.}

\textcolor{black}{Collectively, these considerations underscore the need for focused, resource-aware, and problem-specific approaches in the early fault-tolerant era of quantum computing. They set the stage for exploring chemical opportunities in Sec.~\ref{sec:oppo}, where we examine problem classes and observables that align naturally with the 25--100 logical-qubit regime.}

%%%%%%%%%%%%% Sec.~III

\section{Opportunities in Quantum Chemical Simulations}\label{sec:oppo}

\textcolor{black}{\textit{In this section, we highlight problem classes and observables in quantum chemistry that align naturally with the 25--100 logical-qubit regime, emphasizing cases where quantum solvers can credibly complement or surpass classical approaches.}}

Quantum chemistry presents a set of grand challenges where quantum computers (even at the scale of 25--100 logical qubits) are expected to make meaningful contributions. Rather than aiming for a wholesale replacement of classical electronic structure methods, progress is anticipated through hybrid approaches that integrate quantum computing with classical techniques, high-performance computing (HPC), and artificial intelligence (AI). This synergy enables both targeted enhancement of computational workflows and the exploration of scientific regimes where classical models break down. In the following, we highlight several directions that offer promising opportunities for near-term demonstrations of quantum utility in quantum chemical simulations.

%----------Sec.~III.A

\subsection{Strong Correlation and Active Space Decomposition}

Many chemically and industrially important systems, such as open-shell transition metal complexes~\cite{Leuenberger2001,linert2012molecular,Wasielewski2020} and $f$-electron materials~\cite{Acharya2023,cui2022cuprate,gould2022ultrahard}, exhibit strong electronic correlation. This characteristic poses significant challenges for standard classical simulation methods. Density functional theory, for instance, while widely used in catalysis~\cite{Qin2023Cation}, faces limitations in quantitatively describing strong correlations, electron spin states in magnetic catalysts, and {noncovalent} interactions (dispersion, hydrogen bonding), which can affect barrier predictions. Single-reference wave function approaches also struggle with strong correlation~\cite{Burke2012DFT,Lyakh2012Multireference}. These limitations become particularly pronounced when high precision is essential, such as in modeling intricate catalytic mechanisms where quantitative accuracy is critical for mechanism identification and catalyst design, electronic excitations spanning valence and core levels~\cite{DreuwHeadGordon2005,Souza2024excitedstate}, conical intersections governing photochemical branching, charge-transfer states in chromophores, and relativistic effects in heavy elements~\cite{Pyykko2012relativistic,hess2003relativistic,Autschbach2012Relativistic}. \textcolor{black}{We note that these same considerations underlie the challenge of transition states: near the saddle point, partial bond cleavage induces strong/static (multi-reference) correlation and near-degeneracy that can render single-reference methods unreliable, motivating active-space, downfolded, or embedded treatments targeted along the reaction coordinate.}

\textcolor{black}{This exponential wall is clearly illustrated by the full configuration interaction (FCI) determinant growth,
\begin{align}
%N_{\mathrm{det}} = \binom{N_{\mathrm{orb}}}{N_{\mathrm{elec}}},
N_{\mathrm{det}} = \binom{N_{\mathrm{orb}}}{N_{\rm e}(\alpha)} \times \binom{N_{\mathrm{orb}}}{N_{\rm e}(\beta)}
\end{align}
where $N_{\rm e}(\alpha)$ and $N_{\rm e}(\beta)$ are the numbers of $\alpha$- and $\beta$-electrons, respectively. As can be seen, the growth increases combinatorially with the number of orbitals $N_{\mathrm{orb}}$ and electrons~\cite{doi:10.1021/acs.jctc.3c01190,doi:10.1021/acs.jctc.4c00641}. Even modest active spaces with 20--30 orbitals already lead to billions of determinants, underscoring the limits of brute-force classical treatments. As mentioned in the Introduction, the orbital ranges addressed in recent distributed FCI and relativistic DMRG-tailored CC studies~\cite{doi:10.1021/acs.jctc.3c01190,doi:10.1021/acs.jctc.4c00641}  are comparable to those that can be mapped onto $\sim$25--100 logical qubits under conventional Jordan-Wigner or Bravyi-Kitaev fermion-to-qubit mappings, underscoring why this regime represents the first realistic window for quantum utility (see Sec.~\ref{sec:validation} for representative systems and resource estimates).}

\textcolor{black}{Although the exponential wall ultimately limits all multi-reference methods, modern implementations have substantially extended their practical reach. Tensor network approaches such as DMRG (and its relativistic extensions), stochastic configuration interaction, and quantum Monte Carlo can now simulate thousands of orbitals on leadership-class HPC platforms. Their efficiency is, however, strongly system dependent: tensor networks are limited by entanglement growth and orbital ordering, quantum Monte Carlo by the fermion sign problem, and embedding approaches by fragment–bath partitioning and convergence control. These bottlenecks underscore both the remarkable reach of state-of-the-art classical solvers—often within carefully chosen active spaces—and the complementary potential of quantum algorithms to tackle the remaining hard cases.}  

\textcolor{black}{Consequently, strongly correlated systems are prime candidates for quantum-accelerated solvers~\cite{McArdle2020,Bauer2020,Mazzola2024MonteCarlo} and for advanced embedding strategies~\cite{SunDMETReview2016,RevModPhys.78.865,negre2025newperspectivesdensitymatrixembedding,Ma2021Embedding,Manby2021Embedding,Jones2020Embedding,Libisch2014Embedded,Jacob2024Subsystem,Wesolowski2015Embedding,PhysRevA.109.022418,waldrop2021projector,waldrop2025projector}. Embedding itself spans a spectrum: in some cases the correlation is spatially localized, motivating fragment-based treatments, while in other cases the correlation is delocalized across extended lattices, requiring systematic downfolding approaches (see Sec.~\ref{sec:downfolding} and Deep-Dive~\ref{2}).}  

\textcolor{black}{Electron correlation is not universally local: in strongly delocalized systems such as frustrated magnets or high-$T_c$ superconductors, correlation spans extended lattices and cannot be captured by fragment-based treatments. In such cases, downfolding techniques provide a more appropriate route to derive effective Hamiltonians that retain essential delocalized physics within compact forms. By contrast, in many molecular and bioinorganic systems, correlation effects are spatially concentrated within active sites or fragments. This motivates fragment-based strategies such as the Localized Active Space (LAS) approach, which has emerged as a promising framework for transition-metal complexes, molecular magnets, and bioinorganic clusters~\cite{Pandharkar2021Localized,Agarawal2024Automatic,Pandharkar2022Local,Otten2022Localized}.}

LAS constructs the total wavefunction as an antisymmetrized product of local active space wavefunctions defined on weakly entangled fragments. Each fragment is treated with high-level methods, while inter-fragment interactions are captured at a mean-field level. The LAS State Interaction (LASSI) refinement recovers spin symmetries by diagonalizing the full Hamiltonian within a basis of LAS-configured states. Notably, the inclusion of charge transfer (CT) configurations between fragments within the LASSI Hamiltonian can be crucial for achieving quantitative accuracy, for example, when calculating magnetic coupling constants in multinuclear complexes. 

The applicability and scalability of LAS/LASSI have been demonstrated through calculations on systems like Cr(III) dimers~\cite{Pantazis2019ChromiumDimer,Sharma2020magnetic,Pandharkar2022Local} and spin ladder in Fe$_3$ compound~\cite{Agarawal2024Automatic}, and large-scale LASSCF calculations are being applied to systems with over 1000 orbitals (e.g., Fe$_4$S$_4$, Cr$_2$, NiFe$_2$, and Ni$_2$). LAS is part of a broader family of embedding techniques; related methods like Density Matrix Embedding Theory (DMET)~\cite{Knizia2012DMET} are also being combined with high-level solvers for large systems and specific applications such as core-level spectroscopy~\cite{Jangid2024Core}. Together, these strategies provide 
\textcolor{black}{a promising foundation for integrating classical and quantum workflows, with practical scalability contingent on robust self-consistency~\cite{Otten2022Localized,mitra2024localized,d2024state}. While DMET offers appealing locality and parallelism, we note that single-shot and self-consistent variants can be sensitive to numerical fluctuations in the chemical-potential/correlation-potential update, particularly when the fragment solver is stochastic (e.g., on near-term quantum hardware). In practice, robust outer-loop strategies (e.g., DIIS/Anderson mixing, regularization of the correlation potential, and multi-moment matching beyond density) and reproducibility checks are required to attain stable convergence across fragments.}
\textcolor{black}{At this point it is natural to ask why a quantum computer is needed at all if embedding reduces the problem to a compact active space. While many small active spaces ($\sim$10–20 orbitals) remain tractable for advanced classical solvers, larger chemically relevant fragments (25–50+ orbitals) rapidly encounter exponential or entanglement-driven bottlenecks. We return to this point in Sec.~\ref{sec:validation}, where we connect active-space sizes to representative quantum resource estimates.}

To further reduce complexity, downfolding methods can be employed to derive effective Hamiltonians that retain essential many-body physics within minimal active spaces~\cite{KowalskiSubalgebras2018,KowalskiSubsystemFlow2021,KowalskiSubsystemSelfConsistency2023,BaumanVQE_Downfolded2021,PRXQuantum_4_020313,BaumanExcitedStates2019,KowalskiBaumanDynamics2020,BaumanDownfolding2019,BaumanKowalskiJCP2022,BaumanKowalskiMaterialsTheory2022,Metcalf2020}. 
\textcolor{black}{Chemically motivated diagnostics, such as orbital occupation analysis and entanglement entropy, can provide valuable guidance for selecting reduced spaces~\cite{bauman2024densitymatrixrenormalizationgroup}, but each comes with limitations: occupation numbers neglect two-particle correlation information, and entanglement measures are sensitive to the choice of orbital basis. Similarly, emerging AI-assisted strategies offer promising heuristics but remain an active area of research rather than a solved problem. Nevertheless, these approaches can inform the targeting of chemically significant regions, while highlighting the importance of cross-validation and reproducibility in active-space construction.}

%----------Sec.~III.B

\subsection{Quantum Dynamics and Noise-Informed Simulation}\label{sec:dynamics}

Real-time quantum dynamics, particularly for open quantum systems (OQS), has been highlighted as a promising domain for near-term quantum utility~\cite{Miessen2023QuantumDynamics}. Simulating processes such as photoinduced charge transfer, vibrational energy redistribution, and nonadiabatic transitions provides critical insights into reaction mechanisms and non-equilibrium phenomena beyond static approximations~\cite{Nelson2020nonadiabatic}. However, these simulations are computationally intensive for classical methods. Techniques like Multiconfigurational Time-Dependent Hartree (MCTDH)~\cite{Beck2000MCTDH} and Hierarchical Equations of Motion (HEOM)~\cite{Tanimura1989HEOM} face the ``curse of dimensionality,'' scaling poorly with system size, while tensor network approaches can be limited by the area law in capturing highly entangled dynamics compared to quantum circuits~\cite{PhysRevA.111.032409,PhysRevLett.124.137701}.

Quantum devices may be well-suited to this regime due to their natural ability to implement unitary time evolution and sample from high-dimensional entangled states. Methods such as Trotterized real-time evolution, variational dynamics~\cite{otten2019noise}, and Krylov-subspace propagation (related to quantum signal processing, QSP, or Qubitization) have been proposed as viable quantum algorithms (see e.g. Ref. \citenum{Miessen2023QuantumDynamics} for a recent overview). However, challenges remain, including the accurate preparation of initial states (which can be non-trivial and propagate errors), significant measurement overhead due to wavefunction collapse, and the difficulty of accurate Trotterization, especially for coupled, time-dependent systems. 

Simulating OQS presents the additional challenge that quantum computers naturally perform unitary evolution, while open systems exhibit non-unitary dynamics due to environmental interaction. Standard quantum approaches to OQS include embedding the system within a larger environment simulated unitarily (ancilla-based methods), using stochastic quantum trajectories, or implementing Kraus operators~\cite{Hu2020,Schlimgen2021OQS_PRL,PhysRevResearch.6.023263,Carballeira2021,Dan2025nonMarkovian,Delgado2025Quantum,Scholes2025QIS}. Intriguingly, noise, traditionally viewed as an impediment, has been proposed as a resource~\cite{Harrington2022Dissipation}. Hardware-induced decoherence, if properly characterized or engineered, may serve as a proxy for environmental interactions, thereby facilitating OQS simulation. This ``noise-assisted'' approach has been explored in analog  and digital quantum computing~\cite{Guimaraes2023NoiseAssisted,GarciaPerez2020,Daley2022AnalogAdvantage,PhysRevA.108.062424}, potentially reducing overhead compared to full error mitigation by using techniques like partial error correction~\cite{otten2019accounting,otten2019recovering} or pulse control.
\textcolor{black}{From the perspective of quantum utility, the significance of treating noise as a resource is not that noise can be engineered (classical simulations can emulate noisy channels), but that hardware-induced decoherence can be harnessed \emph{in situ} within quantum workflows. This enables hybrid strategies where noise directly contributes to modeling open-system dynamics at scales beyond tractable classical simulation, linking the presence of hardware imperfections to scientifically meaningful predictions.}

The encoding of bosonic modes, representing vibrational, solvent, or bath degrees of freedom, also requires careful consideration. Strategies include truncated Fock spaces, coherent-state encodings, and squeezed-state representations~\cite{Peng2025}. 
\textcolor{black}{The simplified operator structure in exciton–boson models (often $O(N^2)$ or $O(N)$ terms, versus $O(N^4)$ for generic electronic Hamiltonians) is advantageous for {both} classical and quantum solvers. In the fault-tolerant setting, it also translates directly into lower block-encoding and Trotter (or linear combination of unitaries, LCU) costs (i.e., smaller term sums and $L_1$ norms), fewer oracle calls under qubitization, and simpler state-preparation primitives (e.g., Gaussian/coherent/squeezed states and qumode encodings), thereby reducing the logical-qubit and $T$-count budgets for dynamics simulations. We emphasize that classical methods benefit from this structure as well; however, long-time, finite-temperature, or non-Markovian dynamics remain challenging in practice due to tier growth in HEOM, the curse of dimensionality in MCTDH, and entanglement-growth-limiting tensor networks. By contrast, fault-tolerant quantum simulation avoids the real-time sign problem and can exploit the reduced term count to achieve gate/query complexities that scale with the Hamiltonian norm and simulation time, making exciton–boson platforms promising early targets for resource-aware quantum dynamics~\cite{Tanimura1989HEOM, Beck2000MCTDH, Miessen2023QuantumDynamics, LowHamiltonianSim2017, Berry2019qubitizationof, Low2019hamiltonian}.}
Furthermore, embedding approaches  and the development of advanced Gaussian ansatzes are being explored to reduce the resource requirements for representing these bosonic degrees of freedom~\cite{younas2024spinphonondynamicssinglemolecular}. Non-Hermitian Hamiltonians are also relevant for modeling dissipation, although simulating their dynamics on quantum hardware often involves mapping back to larger unitary systems or specific simulation techniques~\cite{Delgado2025Quantum}. 

%----------Sec.~III.C

\subsection{Hybrid Pipelines and AI Integration}\label{sec:verify}

The integration of quantum computing (QC) with artificial intelligence (AI) and high-performance computing (HPC) is rapidly advancing, with major initiatives developing platform-level solutions for accelerating chemistry and materials discovery~\cite{Alexeev2024AIforQC,vandam2024endtoendquantumsimulationchemical,BECK202411,HoeflerAdvantage2023,klusch2024quantumartificialintelligencebrief,mohseni2024build}. 
\textcolor{black}{AI’s role spans the entire quantum computing stack~\cite{Alexeev2024AIforQC}. In hardware development, it accelerates qubit characterization, architecture exploration, and control pulse optimization. During operation, AI automates calibration and tuning, enabling closed-loop control strategies adaptive to evolving noise environments. In software, AI aids circuit synthesis and compression, variational optimization, and hybrid workload scheduling. It enhances QEC by improving decoder performance~\cite{GoogleLogicalProcessor2024} and enabling scalable, low-latency strategies. In post-processing, AI reduces measurement overhead and mitigates errors in tasks like tomography~\cite{ortegaochoa2025tomographicinterpretationstructurepropertyrelations}, observable estimation, and readout classification. This end-to-end integration of AI is increasingly viewed as essential for making scalable, fault-tolerant quantum computing practical, particularly in the near term. AI contributes by improving decoder throughput for QEC, reducing calibration and scheduling overhead, and lowering measurement costs, which are bottlenecks that otherwise limit the effective utilization of hybrid quantum–classical resources.}

\textcolor{black}{Recent efforts illustrate how industry and academic platforms are beginning to integrate QC, AI, and HPC resources in practice. We cite several examples below only as representative implementations, while the key role of such infrastructures is to enable tight feedback loops, reduce measurement overheads, and improve reproducibility in early fault-tolerant quantum chemistry.}
One effort is NVIDIA’s DGX Quantum system, which enables low-latency, tightly coupled execution between quantum processing units (QPUs) and GPUs to support real-time AI-assisted QEC, calibration, control, and readout~\cite{Alexeev2024AIforQC}. To program such heterogeneous quantum–classical systems, NVIDIA introduced the CUDA-Q platform~\cite{kim2023cuda}, a single-source, hardware-agnostic programming framework that unifies quantum and classical workflows, leveraging NVIDIA's existing CUDA and AI ecosystems.
Another effort is provided by Microsoft’s Discovery platform (formerly, Azure Quantum Elements), which offers a cloud-integrated platform that provides access to multiple quantum hardware backends alongside Azure’s HPC infrastructure and AI toolkits, enabling users to build scalable hybrid quantum applications within a managed cloud environment~\cite{vandam2024endtoendquantumsimulationchemical}.
\textcolor{black}{Complementing these, IBM’s {Qiskit Runtime} provides a cloud-native execution model with low-latency {sessions} and server-side {primitives} (Estimator, Sampler) that streamline hybrid quantum–classical loops and support IBM’s roadmap toward tighter integration of quantum and classical resources for scientific computing~\cite{QiskitRuntime}.}

In such hybrid ecosystems, AI models, which are often trained on large synthetic datasets generated via HPC simulations or augmented with quantum data, are used to accelerate discovery by enabling rapid screening, property prediction, and system-level optimization. For example, AI models deployed within Azure Quantum Elements have been used to evaluate millions of potential battery materials~\cite{Chen2024accelerating}, while NVIDIA has developed generative AI approaches such as GQE~\cite{nakaji2024generativequantumeigensolvergqe} and QAOA-GPT~\cite{tyagin2025qaoa} to assist in synthesizing quantum circuits with features such as reduced depth and enhanced expressivity. In addition to algorithm development, NVIDIA has also demonstrated the use of AI for QEC in collaboration with QuEra~\cite{nvidia_quera_ai_qec}.

This tiered, adaptive computational approach allocates quantum resources to the most challenging subproblems where classical methods struggle, such as the transition-state barrier heights poorly captured by DFT or systems with strong multireference character~\cite{PhysRevA.101.032510,Duan2022}. AI is integral across this hybrid workflow: from hardware design and calibration to device control, algorithm optimization, QEC decoding, and post-processing~\cite{Cao2019,Broughton2021}. Diagnostic tools, including uncertainty quantification in AI models, convergence analysis of classical solvers (e.g., DMRG), and AI-driven screening and circuit optimization, support dynamic resource allocation across the classical-quantum-AI stack~\cite{Bilbrey2025,Moritz2004convergence,Yang2024AI}.

%----------Sec.~III.D

\subsection{Benchmarking and Experimental Validation}\label{sec:validation}

Robust benchmarking remains essential for validating quantum methods and ensuring reproducibility. A three-way validation framework, incorporating quantum simulations, high-level classical reference data, and experimental observations, has been endorsed~\cite{NASQuantumReport2019,NAP26850,Bauer2020}. 
\textcolor{black}{Complementary to these strategies, stabilizer circuits, comprising Clifford-only operations, offer a powerful route for algorithmic validation. Because they can be simulated efficiently on classical hardware~\cite{PhysRevA.70.052328,GoogleSurfaceCode2023,GoogleLogicalProcessor2024,Krinner2022}, they enable cross-checking of quantum compilation, execution, and error correction pipelines at a scale comparable to chemically relevant circuits. Embedding stabilizer-based benchmarks within hybrid workflows thus provides a practical way to verify algorithmic performance in the 25--100 logical-qubit regime before deploying more general, classically intractable simulations. This framework allows for iterative refinement of algorithms and facilitates the identification of performance gaps. }

\textcolor{black}{\textbf{Validation when classical ground truth is unavailable.} For systems that are classically intractable, we advocate a three-tier {validation ladder}~\cite{McArdle2020,Bauer2020,HoeflerAdvantage2023}: (i) {algorithmic self-checks}, enforcing symmetries and conservation laws, variational bounds where applicable, Hellmann–Feynman consistency between energy derivatives and forces, and sum-rule/Kramers–Kronig-type constraints for response functions; (ii) {reduced/embedded cross-checks}, validating predictions on classically tractable fragments or limits of the model (e.g., weak-coupling, non-interacting, or localized-fragment regimes), and bracketing predictions with multiple physically motivated embeddings/downfoldings; (iii) experiment-linked, environment-aware validation, including comparison to observables measured under well-documented conditions, emphasizing differential quantities (e.g., isotope shifts, redox or excitation differences within a series, spectral shifts upon functionalization) that are less sensitive to absolute environment errors.}
\textcolor{black}{For embedding workflows, we recommend reporting embedding-specific convergence diagnostics, such as chemical-potential and density residuals, fragment–bath mismatch, and sensitivity to solver noise, together with variance estimates if the fragment solver is stochastic. When quantum solvers are used, stabilizer-circuit checkpoints provide classically verifiable validation of compilation and execution paths before running classically intractable circuits.}

\textcolor{black}{\textbf{Classical convergence scaffolding.} Many components of hybrid quantum–classical workflows can be benchmarked purely classically prior to any quantum execution and should be used to define, justify, and bound the target instances~\cite{SunDMETReview2016,negre2025newperspectivesdensitymatrixembedding,KowalskiBaumanDynamics2020,BaumanKowalskiJCP2022}. We recommend: (i) {active-space growth} along entanglement/occupation diagnostics (e.g., natural-orbital occupations, single-orbital entropies) with observables plotted vs.\ active-space size; (ii) {embedding region/bath enlargement} with convergence and stability checks (e.g., DMET chemical-potential/correlation-potential residuals, fragment–bath mismatch) under solver noise; (iii) {downfolding/renormalization controls} (commutator truncation order in SES/DUCC, and tensor-factorization ranks swept with observables vs.\ control parameter; and (iv) for {dynamical/open-system models}, bath discretization or tier-depth and mode truncation sweeps to establish time/frequency window fidelity. For each sweep, report stopping tolerances, monotone trends where expected, and confidence intervals; release the raw data/plots as part of the provenance package to support reproducibility. }

\textcolor{black}{To support both capability validation and scientific challenge, we organize benchmarks into infrastructure (Class 0) and challenge (Class 1/2) tracks. {Class 0 benchmarks comprise} chemically grounded but classically tractable sets (e.g., QM9/QH9, MultiXC-QM9, VQM24)~\cite{Yu2023,Nandi2023MultiXC-QM9,Khan2024,McCaskey2019} that enable cross-stack reproducibility, cost-vector calibration, and measurement-reduction studies, often alongside stabilizer-circuit checkpoints. Here, the cost-vector refers to the set of quantitative resource requirements reported for a given simulation, including logical-qubit counts, depth/$T$-count, post-downfolding term counts/$L_1$ norms, and shot budgets needed to achieve target precision. This compact representation enables standardized comparisons across algorithms, hardware platforms, and benchmark instances. {Class 1/2 benchmarks are} also chemically relevant and experimentally tractable, but target computationally demanding instances with multireference character, coupled dynamics, or environment effects (following Ref.~\citenum{morchen2024classificationelectronicstructuresstate}; see also Sec.~\ref{sec:benchmark}). Specifically, we categorize Class 1/2 benchmarks as follows,
\begin{itemize}[]
    \item \textbf{Class 1:} Moderately multireference systems with localized correlations—potentially tractable within 25--100 logical qubits (e.g., cyclobutadiene~\cite{Sierda2023quantum}, Fe$_2$S$_2$ clusters~\cite{Dobrautz2021FeS, Jafari2021RedoxFeS}).
    \item \textbf{Class 2:} Strongly multiconfigurational systems with global entanglement—requiring $>$1000 logical qubits and full fault tolerance (e.g., FeMoco~\cite{Reiher2017}).
\end{itemize}
These systems allow simulations at different levels and experiments to interact closely, {helping guide} method development, hardware requirements, and software stacks across multiple disciplines.}

\textcolor{black}{Table~\ref{tab:combined_benchmarks} reflects this classification, where Class 0 entries are labeled with $\dagger$, and identifies several model systems as focal points for algorithm validation and workflow integration. These span catalysis, photochemistry, and energy materials. For reproducibility, we recommend releasing provenance-rich artifacts (input decks, circuits/Hamiltonians, calibration snapshots, decoder outputs, ensemble definitions, and post-processing scripts) to enable independent audit of environment assumptions and uncertainty budgets. As discussed in Section III.A, while many small active spaces remain tractable with classical solvers, larger chemically meaningful fragments often require quantum resources; the benchmarks here target such cases. To illustrate this connection, Table~\ref{tab:resources} provides resource estimates (logical qubits, circuit depth, and shot counts) for selected exemplar systems, linking orbital counts of practical active spaces to the 25--100 logical-qubit regime. As seen in Table~\ref{tab:resources}, systems such as Fe$_2$S$_2$ clusters or small chromophores map naturally onto the 30--40 logical-qubit range when treated via embedding or downfolding, but still demand circuit depths on the order of $10^4$ and shot budgets exceeding $10^6$. These quantitative estimates highlight the centrality of measurement-efficient ansatzes (Sec.~\ref{sec:ansatze}) and error-aware phase estimation variants (Sec.~\ref{sec:qpe}) to realize practical utility in this regime.}

\textcolor{black}{At the same time, it is important to recognize that even when a solver achieves ``chemical accuracy'' on a model Hamiltonian, this does not necessarily translate into agreement with experiment. Discrepancies can arise from missing environmental effects, conformational sampling, or uncertainties in experimental reference data. Therefore, the community should view chemical accuracy as a baseline threshold, while setting the bar for success in terms of whether quantum simulations can deliver systematic, improvable accuracy for observables of chemical interest (e.g., excitation energies, redox potentials, rate constants). In this context, benchmarking strategies should emphasize not only total energies, but also chemically relevant quantities that can be cross-validated with both experiment and high-level classical methods.}

\textcolor{black}{\textbf{Environment-aware protocol.} When environment effects dominate uncertainty, the validation target should be defined as an {ensemble average} over conformers/solvation states~\cite{RevModPhys.78.865,Jones2020Embedding}. Concretely: (a) specify how the ensemble is generated (e.g., molecular dynamics/Monte Carlo sampling, docking, or curated conformer sets), how weights are assigned (e.g., Boltzmann, reweighting), and how observables are averaged; (b) report a {sensitivity analysis} to environment modeling choices (continuum vs.\ explicit solvent, ionic strength, counterions, pH), and propagate these to an {uncertainty budget} alongside quantum-solver statistical/systematic errors; (c) prioritize {differential} observables and provide {stated error bars} that reflect both ensemble and hardware/runtime contributions. When quantum subroutines (i.e., the quantum-executed portions of a hybrid workflow) are used, include stabilizer-circuit check-points as classically verifiable guards within the workflow.}
\textcolor{black}{For example, while absolute redox potentials of Fe–S clusters in solution or protein environments are far beyond current quantum resources, simplified embedded fragments (e.g., Fe$_2$S$_2$, Fe$_4$S$_4$) can serve as tractable benchmarks, with solvation and environmental contributions incorporated through classical embedding or hierarchical workflows.}\\

\textcolor{black}{In summary, identifying chemically meaningful opportunities and establishing robust validation protocols highlight the need for carefully chosen algorithmic strategies. We turn to these strategies in Sec.~\ref{sec:algo}, focusing on approaches designed to operate effectively within the 25--100 logical-qubit regime.}

\begin{table*}[htbp]
    \centering
    \caption{Representative chemical systems in quantum chemical simulations, including scientific context and simulation goals. \textcolor{black}{Reported accuracy should be interpreted relative to these simulation targets, recognizing that agreement with experiment may further depend on conformational sampling, solvation, and other environmental effects. ${\dagger}$ denotes Class 0 (infrastructure/validation) benchmarks, which are used for implementation verification, cross-stack reproducibility, and measurement-cost studies; not intended as computationally challenging targets. For Class 1/2 (challenge) entries, accompanying materials should include classical convergence baselines (active-space and embedding-region sweeps, downfolding/truncation and tensor-factorization rank studies, and environment-ensemble sensitivity) to justify the specific instances executed on quantum hardware and to support reproducibility.}}
    \label{tab:combined_benchmarks}
    \begin{tabular*}{\textwidth}{p{3.5cm} p{3cm} p{5cm} p{6cm}}
    \hline \hline \\
    $\begin{array}{l}\textbf{Model/System}\end{array}$ & $\begin{array}{l} \textbf{Category}\end{array}$ & $\begin{array}{l}\textbf{Scientific Significance and Features}\end{array}$ & $\begin{array}{l}\textbf{Simulation Target}\end{array}$ \\ \hline \\
    $\begin{array}{l} \text{CO/CO$_2$ on a catalyst} \\ \text{~\cite{Hakkinen2012GoldSulfur,DiPaola2024Platinum,vonBurg2021}} \end{array}$ & $\begin{array}{l} \text{Surface catalysis} \end{array}$ & $\begin{array}{l} \text{Surface-supported reaction network;} \\ \text{Activation barriers} \end{array}$ & $\begin{array}{l} \text{Reaction barrier;} \\ \text{Intermediate species and energetics} \end{array}$  \\ \\
    $\begin{array}{l} \text{Transition metal oxides} \\ \text{~\cite{Pantazis2019ChromiumDimer,Leuenberger2001,Wasielewski2020,Shaik2007Reactivity}} \end{array}$ & $\begin{array}{l} \text{Metal complexes} \end{array}$ & $\begin{array}{l} \text{Strong correlation;} \\ \text{Oxygen atom transfer} \end{array}$ & $\begin{array}{l} \text{Spin states;} \\ \text{Oxidation potentials} \end{array}$ \\ \\
    $\begin{array}{l} \text{Chromophores in solvents}\\ \text{~\cite{Gilmore2008quantum,Fleming1996chromophore,Ji2025entanglement}} \end{array}$ & $\begin{array}{l} \text{Photochemistry} \end{array}$ & $\begin{array}{l} \text{Intramolecular proton transfer;} \\ \text{Fluorescence;} \\ \text{Vibronic and nonadiabatic effects} \end{array}$  & $\begin{array}{l} \text{Excited-state spectra;} \\ \text{Charge/exciton recombination} \end{array}$ \\ \\
    $\begin{array}{l} \text{PCET in enzymes} \\ \text{~\cite{Hammes-Schiffer2009PCET,Schlimgen2021OQS_PRL,Harrington2022Dissipation}} \end{array}$ & $\begin{array}{l} \text{Open quantum} \\ \text{systems} \end{array}$ & $\begin{array}{l} \text{Coupled nuclear-electronic dynamics;} \\ \text{Tunneling and dissipation} \end{array}$ & $\begin{array}{l} \text{Vibronic coupling;} \\ \text{Solvent decoherence} \end{array}$ \\ \\
    $\begin{array}{l} \text{Li[Fe/Mn/Ni/Co]$_y$O$_x$} \\ \text{~\cite{ShokrianZini2023quantumsimulationof,PhysRevA.106.032428,Genreith-Schriever2024}}\end{array}$ & $\begin{array}{l} \text{Battery materials} \end{array}$ & $\begin{array}{l} \text{Redox-driven polymorphism;} \\ \text{Multiconfigurational complexity} \end{array}$ & $\begin{array}{l} \text{Phase transitions and Jahn-Teller effect;} \\ \text{Polaron hopping} \end{array}$ \\ \\
    $\begin{array}{l} \text{Benzene, OLED molecules} \\ \text{~\cite{Sennane2022BenzeneVQE,Plasser2012NewtonX, Sancho-García}} \end{array}$ & $\begin{array}{l} \text{Correlated organics} \end{array}$ & $\begin{array}{l} \text{$\pi$-electron delocalization;} \\ \text{Singlet-Triplet Inversion and gaps} \end{array}$ & $\begin{array}{l} \text{Benchmarking correlated methods;} \\ \text{Gap prediction} \end{array}$ \\ \\
    $\begin{array}{l} \text{Iron-sulfur clusters} \\ \text{~\cite{Dobrautz2021FeS,Jafari2021RedoxFeS}} \end{array}$ & $\begin{array}{l} \text{Strong correlation} \\ \text{testbeds} \end{array}$ & $\begin{array}{l} \text{Magnetic coupling and spin} \\ \text{frustration in bioinorganic settings} \end{array}$ & $\begin{array}{l} \text{Multi-reference solver performance} \end{array}$ \\ \\
    $\begin{array}{l} \text{Alkali metal hydrides} \\ \text{(NaH, KH, RbH)$^{\dagger}$~\cite{McCaskey2019}} \end{array}$ & $\begin{array}{l} \text{Quantum computing} \\ \text{benchmarks} \end{array}$ & $\begin{array}{l} \text{Evaluation of quantum computing} \\ \text{performance for electronic} \\ 
    \text{structure calculations} \end{array}$ & $\begin{array}{l} \text{Ground-state energy calculations} \\ \text{on quantum hardware} \end{array}$ \\ \\
    $\begin{array}{l} \text{QM9 molecules with} \\ \text{QH9, MultiXC-QM9} \\ \text{datasets$^{\dagger}$~\cite{Yu2023,Nandi2023MultiXC-QM9}} \end{array}$ & $\begin{array}{l} \text{Machine learning in} \\ \text{quantum chemistry} \end{array}$ & $\begin{array}{l} \text{Prediction of Hamiltonian matrices} \\ \text{using supervised learning} \end{array}$ & $\begin{array}{l} \text{Accelerated electronic structure} \\ \text{predictions} \end{array}$ \\ \\
    $\begin{array}{l} \text{VQM24 dataset} \\ \text{molecules$^{\dagger}$~\cite{Khan2024}} \end{array}$ & $\begin{array}{l} \text{Large-scale quantum} \\ \text{chemical datasets} \end{array}$ & $\begin{array}{l} \text{Comprehensive coverage of small} \\ \text{molecules for benchmarking} \end{array}$ & $\begin{array}{l} \text{Evaluation of quantum chemical} \\ \text{methods across diverse molecules} \end{array}$ \\ \\
    $\begin{array}{l} \text{Non-equilibrium} \\ \text{non-covalent} \\ \text{complexes~\cite{Sparrow2021}} \end{array}$ & $\begin{array}{l}\text{Non-covalent} \\ \text{interactions} \end{array}$ & $\begin{array}{l} \text{Benchmarking interaction energies} \\ \text{in non-equilibrium geometries} \end{array}$ & $\begin{array}{l} \text{Assessment of computational} \\ \text{methods for noncovalent interactions} \end{array}$ \\ \hline
    \end{tabular*}
\end{table*}

\begin{table*}[t]
\centering
\caption{\textcolor{black}{Representative systems and estimated resource requirements in the 25--100 logical-qubit regime. Estimates are order-of-magnitude, assuming conventional Jordan-Wigner or Bravyi-Kitaev mappings and resource-aware algorithms (downfolding, subspace methods, iterative phase estimation). These examples correspond to chemically motivated active spaces (e.g., Fe--S clusters, chromophores) that are too large for brute-force classical solvers yet fall naturally within the reach of early fault-tolerant quantum devices.}} \label{tab:resources}
\begin{tabular}{llllll}
\hline \hline
&&&&&\\
\textbf{Model/System} & \textbf{Active Space (orbitals)} & \textbf{\# of Logical Qubits} & \textbf{Circuit Depth/$T$-count} && \textbf{\# of Shots} \\
\hline &&&&&\\
{black}
Fe$_2$S$_2$ cluster (bioinorganic model) 
 & $\sim 50$ orbitals (embedded) & 30--50 & $10^4$--$10^5$ && $10^6$--$10^7$ \\
Chromophore (photochemistry) 
 & 20--30 orbitals & 25--30 & $10^3$--$10^4$ && $10^5$--$10^6$ \\
Benzene (aromatic benchmark) 
 & 30 orbitals & 35--45 & $10^4$ && $10^6$ \\
Alkali hydrides (NaH, KH) 
 & 10--12 orbitals & 20--25 & $10^3$ && $10^5$ \\
\hline
\end{tabular}
\end{table*}

%%%%%%%%%%%%%%%%Sec.~IV

\section{Algorithmic Innovations}\label{sec:algo}

\textcolor{black}{\textit{In this section, we survey algorithmic strategies most relevant for the 25--100 logical-qubit regime, focusing on representative methods, their theoretical kernels, resource requirements, and trade-offs with classical approaches.}}

The development of quantum algorithms capable of addressing chemically relevant problems using 25--100 logical qubits must be guided by co-design principles that account for fault-tolerant constraints. These include limited circuit depth, restricted qubit connectivity, noise resilience, and modular architecture~\cite{PRXQuantum.4.030307, zhang2025faulttolerantquantumalgorithmsquantum}. The goal is to create a unified framework for evaluating algorithms across diverse problems, establishing robust metrics and methodologies to uniformly measure performance~\cite{GSEEBenchmarkProject}. Rather than converging on a single dominant paradigm, the current landscape favors the parallel exploration of diverse algorithmic strategies tailored to specific problem structures and hardware capabilities~\cite{Cao2019, McArdle2020, Bauer2020}. Many algorithms exist, balancing accuracy and efficiency, as exact solutions scale exponentially and are often impractical for large systems. 

\begin{figure}
\centering
 \includegraphics[width=\linewidth]{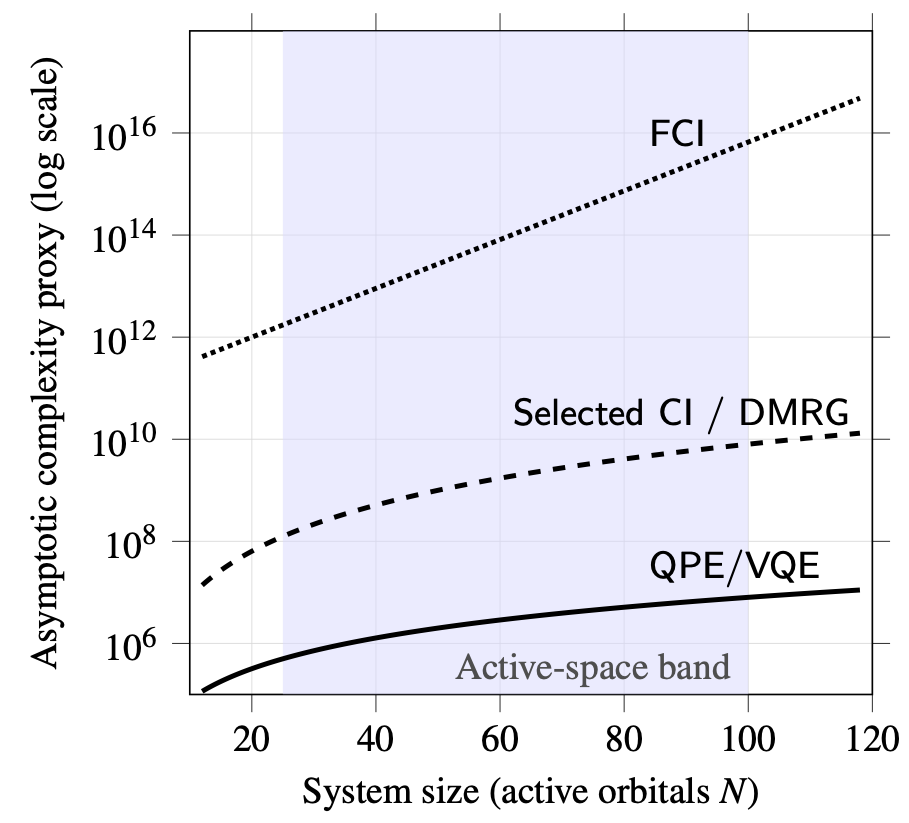}
\caption{\textcolor{black}{Schematic complexity proxies vs. active-space size (orbitals). For FCI, we proxy complexity by determinant growth (exponential/combinatorial), anchored by the trillion-determinant C$_3$H$_8$/STO-3G calculation on 256 servers~\cite{doi:10.1021/acs.jctc.3c01190}. Selected CI/DMRG is shown with polynomial scaling and large prefactors, especially in 4c-relativistic DMRG-tailored CC~\cite{doi:10.1021/acs.jctc.4c00641}, where prefactors approach two orders of magnitude above the non-relativistic case. Quantum (QPE/VQE) uses gate/query complexity after factorization ($\sim\mathcal{O}(N^2)$); the shot budget, Eq.~\eqref{eq:measure}, is reported separately in the text. Normalizations are chosen so that within the 25–100 logical-qubit band (shaded) FCI is $\sim4$ orders above Selected CI/DMRG and the Selected CI/DMRG is $\sim2$ orders above QPE/VQE. Curves are qualitative but anchored to published points and scaling laws.}}
\label{fig:scaling_schematic}
\end{figure}

\textcolor{black}{\textbf{Scope and selection rationale.} The directions highlighted below were selected using criteria tailored to the 25--100 logical-qubit regime: (i) ability to meet tight logical-qubit, depth, and $T$-count budgets (including modest ancilla and code distance); (ii) alignment with chemistry workflows (active-space/embedding, downfolding, and dynamics) with observable-level outputs; (iii) mitigations for noise and trainability (e.g., shallow/subspace circuits, iterative/filtered QPE, error-aware outer loops); (iv) transparent resource models and benchmarkability; and (v) a clear path toward fault-tolerant realization. Methods that currently depend on resources outside this envelope (e.g., deep phase estimation without compression, qRAM-intensive routines) were deprioritized for this near-term discussion. Given known impediments in variational quantum algorithms (VQAs), such as the measurement overhead, barren plateaus, gate-fidelity thresholds, and classical simulability in plateau-free regimes~\cite{VQAChallenges2307,VQAPlateausNPJ2024,VQAClassical2312}, we include only {shallow/subspace} variational approaches paired with measurement-reduction and noise-aware outer loops; hardware-agnostic, deep VQE heuristics are deprioritized in this regime. As summarized in Fig.~\ref{fig:scaling_schematic}, these directions must be understood in light of the contrasting scaling behaviors of classical and quantum approaches. In contrast, resource-aware quantum algorithms such as qubitization-based QPE or measurement-efficient VQE variants target lower-order polynomial gate/query complexity, albeit with explicit shot budgets. Within the 25--100 logical-qubit band, this yields the conservative but realistic ordering that FCI remains several orders of magnitude more costly than Selected CI/DMRG, which in turn is about two orders higher than resource-aware quantum approaches. This hierarchy motivates the structured ansatzes, efficient measurement strategies, downfolding and dimensional-reduction methods, and compressed QPE techniques discussed in the following subsections.}

%----------Sec.~IV.A

\subsection{\textcolor{black}{Structured Ansatzes and Measurement-efficient Subspaces}} \label{sec:ansatze}

\textcolor{black}{\textit{Why included: shallow/subspace constructions offer multi-state access within strict depth/$T$ budgets, integrate with downfolded or embedded active spaces, and can be paired with measurement-reduction (e.g., shadows/overlap estimators) and noise-aware outer loops for trainability.}}

\textcolor{black}{Variational quantum algorithms (VQAs), including VQE~\cite{Peruzzo2014}, face several well-documented challenges. Among these, shot noise and measurement overhead are often cited as particularly demanding in practice, since sample requirements scale with Hamiltonian variance and target precision. However, even in noise-free emulators, VQE remains a heuristic constrained optimization without guaranteed convergence, and practical performance depends strongly on the ansatz, optimizer, and problem structure~\cite{Cerezo2021}. 
These limitations, together with barren plateaus, spin contamination, and hardware noise, underscore why structured ansatzes and measurement-efficient strategies are essential. 
At the same time, rapid progress in hardware~\cite{Ezratty2024UQT,Preskill2025Megaquop,osti_2588210} provides a dynamic backdrop in which these challenges must be continually re-evaluated. While neural network ansatzes have been explored, particularly for solids and lattice models~\cite{PhysRevB.107.235139}, they are not chemically intuitive and remain less common in quantum chemistry. By contrast, most innovation for molecular applications has focused on chemically inspired and hardware-efficient ansatzes~\cite{Kandala2017Nature,fedorov2022unitary}. We have identified four major directions below.}

\textcolor{black}{First, different parameterizations of the unitary Cluster Jastrow (uCJ) ansatz~\cite{Matsuzawa2020Jastrow,MHG_uCJ_details} have been explored. In the original formulation~\cite{Matsuzawa2020Jastrow}, the orbital-rotation generator $K$ is a complex anti-Hermitian matrix, which in practice is often restricted to a real antisymmetric form, while the Jastrow matrices $J$ are pure imaginary and symmetric.  This “real $K$ / imaginary $J$” restriction is the most widely used, as it yields a compact, chemically meaningful parameterization that captures strong correlation more effectively than generalized unitary coupled cluster with singles and doubles (GUCCSD)~\cite{Lee2019GUCC}, while significantly reducing quantum resource requirements~\cite{MHG_uCJ_details}.} Alternative parameterizations such as the imaginary symmetric (Im-uCJ) and general complex (g-uCJ) forms are also promising, particularly for bond dissociation. These findings suggest that tuning the parameterization strategy can improve accuracy and robustness, though care must be taken to evaluate potential drawbacks such as spin contamination. 
\textcolor{black}{We treat adaptive/heuristic enhancements (e.g., operator pooling, layer-wise updates) as optional refinements, contingent on reporting trainability diagnostics and {measurement budgets} for target observables. The measurement cost of VQE approaches can be quantified as
\begin{align}
M \sim \mathcal{O}\!\left(\frac{\mathrm{Var}[H]}{\epsilon^2}\right), \label{eq:measure}
\end{align}
where $M$ is the number of measurements required, $\mathrm{Var}[H]$ is the variance of the Hamiltonian, and $\epsilon$ is the target precision. This scaling highlights why measurement-reduction strategies (e.g., classical shadows, low-rank factorizations, grouping of commuting terms) are critical for pushing VQE into the 25--100 logical-qubit regime.}

Second, optimization-free and adjustable subspace construction methods provide versatile alternatives. These include density-matrix-based quantum subspace diagonalization approaches such as quantum subspace expansion (QSE)~\cite{mcclean2017hybrid}, as well as Krylov/Lanczos methods~\cite{PhysRevA.105.022417,Motta_2024,Tkachenko:2024aa,Berthusen:2024aa}, non-orthogonal configuration interaction schemes~\cite{PRXQuantum.4.030307}, quantum computed moment techniques~\cite{Aulicino2022State}, and generator coordinate-inspired strategies~\cite{qugcm}. The subspace basis can range from optimized determinants (e.g., Hartree–Fock states) to coherent states, enabling access to multiple eigenstates while reducing circuit depth and enhancing sampling efficiency. Error mitigation techniques such as shadow tomography~\cite{Huang2020Shadows,Huang2021shadows} may further reduce measurement overhead.
\textcolor{black}{Representative resource estimates and trade-offs for these methods are illustrated in Deep-Dive~\ref{1}.}

Third, adaptive, parallel, and heuristic enhancements to VQE aim to address trainability and efficiency. Adaptive strategies such as ADAPT-VQE~\cite{grimsley2019adaptive,zhu2022adaptive} and operator pooling dynamically construct compact ansatzes tailored to the correlation structure, minimizing depth without sacrificing accuracy. Parallel parameter optimization has been introduced to accelerate convergence~\cite{stein2022eqc, ParallelVQE2025}, and heuristic methods inspired by quantum annealing have been applied to improve state preparation and guide optimization~\cite{PhilipVQEHeuristic2023}. Complementary approaches like PermVQE~\cite{Tkachenko:2021aa} and ClusterVQE~\cite{Zhang:2022aa} use entanglement-informed qubit permutations or graph-based decompositions to reduce depth and enable scalable parallelization.

Finally, automation and user-accessible deployment of quantum solvers are gaining traction. Recent efforts focus on automating ansatz and subspace construction to facilitate broader adoption of non-orthogonal quantum solvers~\cite{Zheng2024GCM}, thereby reducing the manual complexity of circuit design, subspace definition, and error mitigation integration. Such developments aim to make quantum solvers accessible to a wider community beyond quantum algorithm specialists.
\textcolor{black}{For clarity, the main families discussed above are organized in Table~\ref{tab:ansatz_summary}, which summarizes kernels, resource profiles, and trade-offs in the 25--100 logical-qubit regime.}

\textcolor{black}{Taken together, these structured ansatzes and subspace strategies illustrate the trade-off between expressivity and resource efficiency. The common theme across uCJ, Krylov/subspace, and adaptive refinements is to leverage chemically motivated structure, shallow depth, and measurement reduction to achieve tractable scaling under 25--100 logical-qubit budgets. These methods serve as modular building blocks for embedding and downfolding frameworks (Sec.~\ref{sec:downfolding}), where compact effective Hamiltonians can be paired with such solvers.}

% --- Deep-dive: Subspace and Krylov Methods ---
\begin{DeepDive}[]{Subspace and Krylov Methods}{1}
\textcolor{black}{
\begin{itemize}[leftmargin=1.2em]
\item \textbf{Kernel.} Construct low-depth basis states (e.g., determinants, Krylov vectors) and diagonalize the Hamiltonian projected into this subspace. Quantum subspace expansion and Krylov methods provide multi-state access with shallow circuits~\cite{mcclean2017hybrid,PhysRevA.105.022417,Motta_2024,RobledoMoreno2025chemistry,yu2025quantumcentricalgorithmsamplebasedkrylov}.
\item \textbf{25--100 LQ resource vector.} \begin{itemize}[leftmargin=1.2em] \item \textit{Qubits:} 30--80 (depending on active space size). \item \textit{Circuit Depth/$T$-count:} $\sim 10^3$ per basis vector. \item \textit{Shots:} $10^5$--$10^6$ per observable, reducible by shadows/grouping. \end{itemize}
\item \textbf{Features.}
  \begin{itemize}[leftmargin=1.2em]
    \item Provide multi-state capability with shallow circuits.
    \item Compatible with measurement-reduction strategies and error-mitigation techniques.
    \item Accuracy and cost depend on basis selection and subspace size.
  \end{itemize}
\item \textbf{Best vs.} Standard VQE when multiple excited states or spectra are needed within tight qubit budgets.
\end{itemize}}
\end{DeepDive}

\begin{table*}[t]
\centering
\caption{\textcolor{black}{Representative structured ansatzes and heuristic refinements in the 25--100 logical-qubit regime.}}
\label{tab:ansatz_summary}
%\resizebox{\textwidth}{!}{%
\begin{tabular}{lllll}
\hline \hline
&&&& \\
\textbf{Method} && \textbf{Key Idea / Kernel} && \textbf{Notes for 25--100 logical-qubit regime} \\
\hline
&&&&\\
\makecell[l]{uCJ family \\ (real/imag/g-uCJ)} && \makecell[l]{Compactly capture correlation; \\ 
Robust across bond breaking~\cite{Matsuzawa2020Jastrow,Lee2019GUCC}.} && \makecell[l]{25--60 logical qubits; Depth $\sim 10^3$--$10^4$; \\ Compact, spin-adaptable; \\ Trainability depends on parameterization.} \\
&&&&\\
\makecell[l]{ADAPT-VQE and \\ automation tools}  && \makecell[l]{Iteratively grows ansatz from an operator \\ pool tailored to correlation structure~\cite{stein2022eqc}. \\ Automated ansatz/subspace generation \\ to lower user burden~\cite{Zheng2024GCM}.} && \makecell[l]{Flexible, accessible, and compact; \\ Overhead from quantum-classical loop, \\ sensitive to solver noise.} \\
&&&&\\
Heuristic refinements && \makecell[l]{Operator pooling, parallel updates, \\ annealing-inspired heuristics~\cite{ParallelVQE2025,PhilipVQEHeuristic2023}.} && \makecell[l]{Reduce depth or accelerate convergence; \\ Stochastic noise may destabilize.} \\
\hline
\end{tabular}
%}
\end{table*}

%----------Sec.~IV.B

\subsection{Downfolding and Renormalization Techniques}\label{sec:downfolding}

\textcolor{black}{\textit{Why included: downfolding compresses correlated physics into compact active Hamiltonians that fit 25--100 logical qubits and couple naturally to shallow/subspace solvers.}}

Effective Hamiltonian construction via downfolding offers a powerful way to reduce resource requirements while maintaining chemical accuracy~\cite{BaumanKowalskiMaterialsTheory2022}. By integrating out external degrees of freedom and targeting compact active spaces, these approaches bridge high-level accuracy with qubit efficiency.

One important direction is coupled-cluster–based downfolding, where techniques such as subsystem embedding subalgebras (SES)\cite{KowalskiSubalgebras2018, BaumanDownfolding2019} and double unitary coupled-cluster (DUCC) downfolding\cite{BaumanVQE_Downfolded2021, Metcalf2020} construct effective Hamiltonians within reduced active spaces. The accuracy of these approaches depends critically on the treatment of commutator terms~\cite{BaumanKowalskiJCP2022}, and they have been successfully applied to ground~\cite{BaumanDownfolding2019} and excited states~\cite{BaumanExcitedStates2019} as well as nonequilibrium dynamics~\cite{KowalskiBaumanDynamics2020}. 
%
%\textcolor{black}{Briefly speaking, SES restricts the cluster operator to excitation subalgebras and DUCC uses unitary CC/BCH expansions to integrate out external space, yielding an active-space \(H_{\mathrm{eff}}\) with smaller term norms and counts, thereby reducing logical-qubit and \(T\)-gate budgets at the cost of a controllable commutator truncation. Representative reporting fields and trade-offs for SES/DUCC are summarized in Deep-Dive~\ref{2}.}
\textcolor{black}{In brief, SES- and DUCC-based downfolding provide many-body algorithms that integrate out the so-called external degrees of freedom, yielding an active-space effective Hamiltonian $H_{\rm eff}$
 with reduced term norms and counts, thereby reducing logical-qubit and \(T\)-gate budgets at the cost of a controllable commutator truncation. Representative reporting fields and trade-offs for SES/DUCC are summarized in Deep-Dive~\ref{2}.}

% --- Deep-dive: SES/DUCC downfolding  ---
\begin{DeepDive}[]{SES/DUCC Downfolding}{2}
\textcolor{black}{
\begin{itemize}[leftmargin=1.2em]
  \item \textbf{Kernel.} Construct an effective Hamiltonian on the active space,
  $H_{\mathrm{eff}} = P\, e^{-\sigma_{\mathrm{ext}}} H\, e^{\sigma_{\mathrm{ext}}} P$, where external excitations are integrated out via a truncated BCH series (SES/DUCC). The truncation order and excitation rank in $\sigma_{\mathrm{ext}}$ are explicit accuracy knobs that control both term counts and the $L_1$ norm, directly impacting logical circuit depth/$T$-count for subsequent solvers~\cite{KowalskiSubalgebras2018,BaumanDownfolding2019,BaumanKowalskiJCP2022,BaumanVQE_Downfolded2021,Metcalf2020}.
  \item \textbf{25--100 LQ resource vector.} 
	\begin{itemize}[leftmargin=1.2em]
  	\item \textit{Qubits:} adjustable to $\sim$25--100 logical qubits by tuning the downfolded active space; the upper limit is set by the feasibility of classical preprocessing (cluster amplitudes, commutator evaluation, diagnostics).
  	\item \textit{Circuit depth / $T$-count:} scales with the number of terms in the downfolded Hamiltonian; reduced relative to unreduced Hamiltonians, proportional to post-downfolding term counts or $L_1$ norms.
  	\item \textit{Reporting:} (i) active-space size, (ii) post-downfolding term count and/or $L_1$ norm, (iii) commutator truncation order, and (iv) targeted observable/precision. This makes circuit depth/$T$-count and shot budgets transparent for early-FTQC demonstrations.
	\end{itemize}
  \item \textbf{Features.}
  \begin{itemize}[leftmargin=1.2em]
    \item Compression aligned with chemical intuition.
    \item Explicit approximation knob via commutator truncation order and excitation rank.
    \item Provides a framework for hybrid computing infrastructure that integrates exascale architectures with quantum hardware.
    \item Integrates with multi-component and multi-scale workflows and shallow solvers.
    \item Requires explicit reporting of truncation order and convergence criteria for reproducibility.
  \end{itemize}
\item \textbf{Best vs.} Direct encodings and fragment-only approaches when strong correlation is delocalized: downfold to a compact active $H_{\mathrm{eff}}$ and solve with subspace/QPE variants within 25--100 logical-qubit budgets.
\end{itemize}
}
\end{DeepDive}

Another family of approaches is based on renormalization flow transformations, including quantum flow (Q-Flow) methods, which recast the  energy optimization  in large space into coupled, energy-dependent effective models~\cite{KowalskiSubsystemFlow2021}. 
\textcolor{black}{Notably, the continuous unitary/flow transforms decouple active and external sectors and can encode energy dependence (self-energy-like effects) needed for spectra/dynamics while keeping the quantum register compact.}
Complementary progress has been achieved through tensor factorization techniques, where recursive downfolding leverages tensor decompositions to optimize scaling complexity on both classical and quantum architectures~\cite{BanerjeeTensorDownfold2023}. 
\textcolor{black}{Here, factorizing two-electron tensors reduces memory and effective term counts, lowering LCU/qubitization or Trotter costs and easing Hamiltonian-simulation or subspace-diagonalization on small logical registers.}

Finally, hybrid Green’s function and wave function embedding schemes combine the strengths of both frameworks~\cite{Dvorak2019,Romanova2023}. In such methods, the Green’s function formalism is employed to describe the bath (environment), while explicit wave function solvers treat the active space, yielding an embedding that differs conceptually from purely wave function based approaches.  
\textcolor{black}{Explicitly, the bath is derived from a  Green’s function that captures dynamical environment effects, while the quantum subroutine is confined to the active space for ground/excited states and spectra with explicit resource savings.  A primary advantage of this approach is that Green's functions can computed efficiently and yield a fairly high fidelity electronic structure for the bath.  The challenge is to find a prescription that makes the downfolded active space hamiltonian energy-independent.}

Together, these downfolding and renormalization techniques enable scalable embedding schemes~\cite{SunDMETReview2016, Manby2021Embedding,waldrop2025projector}, remain compatible with both classical solvers and quantum subspace diagonalization, and, when combined with Green’s function embedding and self-energy projection, provide access to spectral and dynamical properties beyond ground states~\cite{Aryasetiawan2022}. 
\textcolor{black}{For reporting, we recommend annotating active-space size, post-downfolding term counts or \(L_1\) norms, commutator truncation order (if applicable), and targeted observables.}

\textcolor{black}{In practice, these compact effective Hamiltonians pair naturally with the subspace and phase-estimation strategies of Sec.~\ref{sec:qpe}, where resource vectors can be stated transparently against post-downfolding $L_1$ and target precision.}

%----------Sec.~IV.C

\subsection{Quantum Phase Estimation (QPE) and Variants} \label{sec:qpe}

\textcolor{black}{\textit{Why included: iterative/filtered/statistical QPE variants retain gold-standard precision while minimizing ancilla and depth, making phase estimation viable in early fault-tolerant settings.}}

QPE remains the gold standard for precision energy estimation~\cite{AspuruGuzik2005}, serving as a key subroutine in many quantum algorithms~\cite{NielsenChuangBook,McArdle2020,Bauer2020}, but its resource demands typically exceed what is available on near-term fault-tolerant devices. To address these limitations, several variants have been introduced to reduce depth, ancilla requirements, and error sensitivity.

Iterative and Bayesian protocols minimize the number of ancilla qubits and circuit depth by trading these resources for more measurements~\cite{PhysRevA.109.032606,PhysRevA.76.030306,YamamotoBayesianQPE2024}. Bayesian approaches in particular optimize parameter selection, improving efficiency and robustness against noise. A related class of methods, known as statistical phase estimation, employs lower-depth circuits and fewer auxiliary qubits, making them more compatible with early fault-tolerant hardware and error-mitigation strategies~\cite{BluntStatQPE2023}. These statistical approaches can achieve higher accuracy for a given circuit depth than earlier analyses.
\textcolor{black}{Representative resource requirements and trade-offs for iterative and filtered QPE are summarized in Deep-Dive~\ref{3}.}

% --- Deep-dive: Iterative and Filtered QPE  ---
\begin{DeepDive}[]{Iterative and Filtered QPE}{3}
\textcolor{black}{
\begin{itemize}[leftmargin=1.2em]
  \item \textbf{Kernel.} Given a block-encoding/qubitization of $H$ with scale $\lambda$, define a walk operator $W$ whose eigenphases satisfy $\cos\theta = E/\lambda$. Phase estimation on $W$ recovers $E=\lambda\cos\theta$ with precision $\epsilon$ using a QSP polynomial of degree $d=\mathcal{O}(\lambda/\epsilon)$. Iterative and Bayesian QPE minimize ancilla width by trading depth for repeated measurements, while Gaussian/subspace filters improve precision under noise~\cite{PhysRevA.109.032606,PhysRevA.76.030306,YamamotoBayesianQPE2024,BluntStatQPE2023,PhysRevA.104.062435}.
  \item \textbf{25--100 LQ resource vector.}
    \begin{itemize}[leftmargin=1.2em]
      \item \textit{Qubits:} 25--100 logical qubits post-downfolding, typically with 1–3 ancillas.
      \item \textcolor{black}{\textit{Circuit Depth/$T$-count:} $10^4$–$10^5$, depending on target precision and filter degree.}
      \item \textcolor{black}{\textit{Shots:} $10^5$–$10^6$ per observable, amortized across iterative/Bayesian schedules.}
    \end{itemize}
  \item \textcolor{black}{\textbf{Features.}}
    \textcolor{black}{
    \begin{itemize}[leftmargin=1.2em]
      \item Provides certificate-grade eigenvalues with systematic error bounds.
      \item Natural access to excited states through phase registers.
      \item Filtered variants mitigate noise sensitivity and improve robustness.
      \item Resource costs scale transparently with post-downfolding $L_1$ norms and target precision.
    \end{itemize}
    }
  \item \textcolor{black}{\textbf{Best vs.} Variational/subspace methods when precision eigenvalues or sharp spectral features are required, within the depth budgets of early fault-tolerant devices.}
\end{itemize}
}
\end{DeepDive}

Together, these QPE variants improve scalability, particularly when combined with classical compression and error-mitigation techniques such as classical shadows~\cite{CaiErrorMitigationReview2023, BluntStatQPE2023}. Their feasibility is underscored by recent demonstrations, including Bayesian QPE on trapped-ion systems~\cite{YamamotoBayesianQPE2024} and statistical phase estimation on superconducting processors~\cite{BluntStatQPE2023}, as well as hybrid QPE–VQE workflows that prepare approximate eigenstates variationally and refine them with phase estimation~\cite{PhysRevLett.122.140504}. Looking ahead, fault-tolerant algorithms based on QPE are also being extended to new observables, such as interaction energies within symmetry-adapted perturbation theory~\cite{CortesSAPT2024}.

%----------Sec.~IV.D

\subsection{\textcolor{black}{Alternative paradigms beyond shallow VQAs}}\label{sec:altparadigms}

\textcolor{black}{\textit{Why included: these approaches target chemistry-relevant observables while mitigating VQA scaling limits and offering transparent resource models.}}

\textcolor{black}{In a {measurement-assisted classical} strategy, classical electronic-structure solvers are {supplemented} by overlap or transition measurements on a quantum device to refine subspaces or non-orthogonal expansions while keeping the quantum footprint small~\cite{JCTC2024Overlap}. {Quantum-assisted Monte Carlo} takes a complementary path: quantum subroutines can {unbias} or accelerate fermionic Monte Carlo by preparing/sign-correcting states or estimating non-classical projectors, with resource–accuracy tradeoffs that are readily benchmarked~\cite{HugginsNature2022}. A related {Hamiltonian-moment hybrid} approach, derived from imaginary-time evolution, leverages quantum-measured Hamiltonian moments to construct systematic classical expansions, offering a compact and improvable route to ground and excited states~\cite{kowalski2020quantum,Vallury2020quantumcomputed,Claudino_2021,seki2021quantum,Aulicino2022State}. For dynamics, {real-time electron-dynamics} schemes provide direct access to time-dependent observables and spectra; the favorable Hamiltonian structure discussed in Sec.~\ref{sec:dynamics} and qubit-efficient encodings help reduce budgets relative to generic electronic Hamiltonians~\cite{NatComm2023ExactDynamics,PRX2023ClassicalOscillators}. Across these paradigms, resources can be stated explicitly, including shot budgets, oracle/query counts, and logical depth/$T$ counts, and typical instances fit within the 25–100 logical-qubit regime while mapping cleanly to the benchmark and validation program outlined in Sec.~\ref{sec:class} and Sec.~\ref{sec:validation}.}

\textcolor{black}{These approaches exemplify how non-VQA strategies can be integrated into the broader benchmark and validation framework developed in Secs.~\ref{sec:class} and~\ref{sec:validation}.}

%----------Sec.~IV.E

\subsection{\textcolor{black}{Benchmarking, Resource Modeling, and Modular Execution}}\label{sec:benchmark}

\textcolor{black}{\textit{Why included: utility claims require standardized resource models, comparable benchmarks, and modular execution to make progress auditable across platforms~\cite{GSEEBenchmarkProject,McCaskey2019}.}}

\textcolor{black}{Realistic resource estimation and apples-to-apples comparisons hinge on reporting a common {cost-vector} for each result: ({i}) logical-qubit count and logical depth/$T$-count (and ancilla/code-distance assumptions); ({ii}) post-downfolding Hamiltonian size (e.g., term counts or $L_1$ norms) and any tensor factorizations; ({iii}) oracle/query counts where applicable; and ({iv}) {shot budgets per observable} (or per target error bar) together with any measurement-reduction used~\cite{GSEEBenchmarkProject,KumarVQEBenchmarking2023,HuangKernel2021}. These costs should accompany the benchmark classes in Sec.~\ref{sec:class} and the validation protocol in Sec.~\ref{sec:validation}.}

\textcolor{black}{To make results reproducible and analyzable, we favor {modular benchmark execution}, e.g., the ground state energy estimate staged flow (problem database $\rightarrow$ solution generation $\rightarrow$ feature computation $\rightarrow$ performance analysis $\rightarrow$ reporting) with unique identifiers and provenance for each stage~\cite{GSEEBenchmarkProject}. Packaging artifacts (inputs, circuits/Hamiltonians, calibration/decoder outputs, feature vectors, and post-processing scripts) enables re-execution and independent audit~\cite{claudino2022xacc,BECK202411}. Difficulty–feature maps and active-learning curation can be used to position and evolve suites~\cite{GSEEBenchmarkProject,Alexeev2024AIforQC,McCaskey2019}, while community programs/toolkits provide standardized harnesses for evaluation~\cite{DARPA_QB_Program,PollardBenchQC2025}; here we focus on the minimal resource/packaging contract needed to make cross-platform comparisons fair and durable. We also encourage depositing these artifacts in a shared hardware-experiment registry (Sec.~\ref{sec:codesign}) to enable cross-platform reuse and AI-driven analysis.}\\

\textcolor{black}{These choices do not preclude other promising directions; rather, they reflect near-term feasibility under the 25--100 logical-qubit constraints and emphasize methods with measurable, benchmarked progress toward chemically validated observables. The next step is to consider how these algorithmic ingredients interact with hardware and classical infrastructure, a topic we explore in Sec.~\ref{sec:pathways} on hybrid integration pathways.}

%%%%%%%%%%%%%%%%Sec.~V
\begin{figure}
    \centering
    \includegraphics[width=0.9\linewidth]{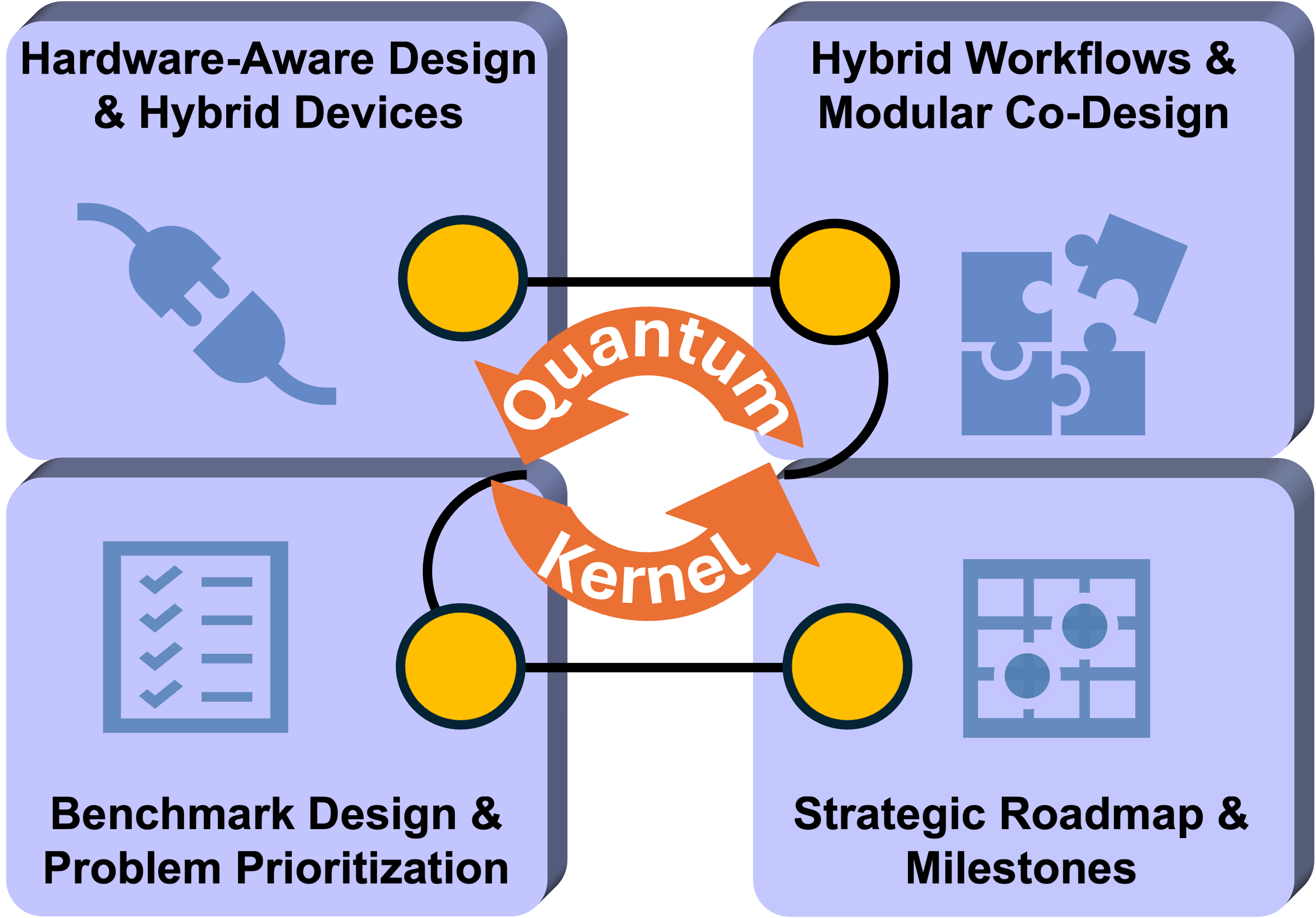}
    \caption{\textcolor{black}{Hybrid workflow schematic aligning with Sec.~\ref{sec:pathways}. The central {Quantum Kernel} connects to four enabling areas: (i) hardware-aware design and hybrid devices, (ii) hybrid workflows and modular co-design, (iii) benchmark design and problem prioritization, and (iv) strategic roadmap and milestones.  This structure illustrates how execution-level workflows and co-design infrastructure jointly enable scalable quantum chemistry simulations in the 25--100 logical-qubit regime.}}
    \label{fig:hybrid_workflow}
\end{figure}

\section{Hybrid Integration Pathways}\label{sec:pathways}

\textcolor{black}{\textit{In this section, we discuss hybrid integration pathways that combine quantum algorithms, classical resources, and AI-assisted infrastructure, showing how these elements can be co-designed to deliver auditable progress within the 25--100 logical-qubit window.}}

Realizing practical quantum simulations in the 25--100 logical-qubit regime necessitates deep integration across quantum algorithms, classical computing resources, and emerging hardware architectures~\cite{vandam2024endtoendquantumsimulationchemical}. Quantum utility in this intermediate scale is unlikely to be achieved by quantum processors in isolation but is increasingly viewed as contingent on well-orchestrated hybrid systems that enable co-designed, end-to-end workflows~\cite{HoeflerAdvantage2023, BECK202411}. Understanding the interplay between algorithms, software stacks, and hardware is paramount.
\textcolor{black}{Figure~\ref{fig:hybrid_workflow} provides an overview of these hybrid integration pathways, highlighting four enabling areas that connect to the central quantum kernel: hardware-aware design and hybrid devices, hybrid workflows and modular co-design, benchmark design and problem prioritization, and strategic roadmap and milestones. This schematic mirrors the organization of the subsections below and emphasizes that both execution-level workflows and supporting co-design infrastructure are required to realize scalable quantum chemistry simulations in the 25--100 logical-qubit regime.}

%----------Sec.~V.A

\subsection{Hardware-Aware Design and Hybrid Devices}\label{sec:hardware}

\textcolor{black}{\textit{Scope:} Algorithm design that accounts for hardware realities (gate fidelities, error rates, connectivity, and QEC throughput) while also leveraging extended resources such as qumodes and hybrid qubit–boson architectures.}

\textcolor{black}{Quantum chemistry workloads place stringent demands on hardware, both in conventional qubit-based architectures and in emerging hybrid modalities such as qubit–qumode devices.}
These demands include high-fidelity two-qubit operations, sufficient error-correction throughput, and fast classical control and I/O. Emerging heterogeneous platforms (e.g., tightly coupled GPU–QPU systems with low-latency interconnects) aim to address these requirements and to host AI-assisted decoding and real-time calibration/control loops~\cite{NVIDIADGXQuantum,Alexeev2024AIforQC}. Practical algorithm design in the 25–100 logical-qubit regime therefore begins with explicit budgets for gate speeds, fidelities, error rates, connectivity, and coherence across leading modalities (superconducting, trapped-ion, atom-array, photonic), together with transpiler flows that account for native gate sets and differential costs (e.g., $R_Z$ vs.\ $\mathrm{CX}$) in the fault-tolerant setting~\cite{wang2024optimizingftqcprogramsqec}. In parallel, heterogeneous QEC architectures that combine surface codes with LDPC-style constructions (e.g., Gross code) are being explored to optimize space–time overheads~\cite{HetEC}.

Three enablers are especially relevant for chemistry: (i) compact fermionic encodings (e.g., Bravyi–Kitaev, parity) to minimize ancilla overhead; (ii) mid-circuit measurement with layout-aware transpilation to shorten depth under connectivity constraints; and (iii) noise-informed scheduling with AI-based decoders, ideally backed by high-fidelity noise models developed with hardware providers~\cite{CaiErrorMitigationReview2023}.

\textcolor{black}{Beyond qubit-only considerations, extended resources such as qumodes (quantum harmonic oscillators) provide complementary opportunities.} 
Qumodes offer an effectively infinite-dimensional Hilbert space and can act as qudits under energy cutoffs~\cite{Weedbrook2012,Wang2020Qudits}. 
In the circuit quantum electrodynamics (cQED) platforms, microwave cavities dispersively coupled to transmons have demonstrated high-quality, long-lived qumodes, universal gate sets, and even break-even quantum error correction for logical qubits and qudits~\cite{Blais2021,Sivak2023real,Brock2024quantum}. 
Such hybrid qubit–qumode architectures are difficult to mimic with shallow qubit-only circuits and are promising for problems with strong bosonic character, including vibronic structure and dynamics~\cite{Wang2020vibronic,Liu2024review}. 
Remaining challenges include mitigating noise from the ancillary qubit coupled to the cavity~\cite{Reagor2016QED}. 
\textcolor{black}{Further details on specific cQED gate implementations and application domains are provided in Appendix~\ref{app:qumode}.}

\textcolor{black}{Delivering these capabilities portably requires co-designed software/hardware interfaces and shared infrastructure. Compiler stacks must coordinate QPU, GPU, and CPU tasks under sub-$\mu$s latencies and bridge the quantum–classical interface using chemistry-aware abstractions and pre-compiled primitives (e.g., Trotter steps, block encodings, qubitization segments)~\cite{LowHamiltonianSim2017,Berry2019qubitizationof,Low2019hamiltonian}. With realistic hardware budgets and transpiler costs in hand, we turn next to the hybrid execution model that concentrates quantum resources where they matter most.}

%----------Sec.~V.B

\subsection{\textcolor{black}{Hybrid Workflows and Modular Co-Design}}\label{sec:codesign}

\textcolor{black}{\textit{Scope:} Hybrid pipelines integrating HPC, quantum kernels, and AI, together with reproducible co-design infrastructure, enabling scalable execution and benchmarking in the 25--100 logical-qubit regime.}

\textcolor{black}{To illustrate the interplay of classical HPC, quantum subroutines, and AI integration, Fig.~\ref{fig:hybrid_workflow} depicts a schematic hybrid workflow.}
The hybrid model, usually referring to classical pre-/post-processing wrapped around a quantum subroutine solving a reduced subproblem, remains the most feasible near-term strategy. In chemistry, quantum acceleration is typically focused on compact active subsystems (multiconfigurational fragments or correlated sites) identified by diagnostics such as DMRG-based entropy measures and orbital entanglement, or via automated schemes like AutoCAS~\cite{Moritz2004convergence,Stein_Reiher_2017,stein2016automated,Stein2016balance,stein2019autocas}. Problem partitioning relies on quantum embedding\cite{SunDMETReview2016,Ma2021Embedding,Manby2021Embedding,Wesolowski2015Embedding,Jacob2024Subsystem,negre2025newperspectivesdensitymatrixembedding,waldrop2021projector,waldrop2025projector} and downfolding\cite{BaumanKowalskiMaterialsTheory2022,PRXQuantum_4_020313,BaumanDownfolding2019,KowalskiSubsystemFlow2021,BanerjeeTensorDownfold2023}, with Green’s-function techniques supplying high-fidelity baths for environment coupling when spectral properties or open-system effects are central~\cite{Aryasetiawan2022,RevModPhys.78.865}; 
\textcolor{black}{and when embedding loops wrap stochastic (quantum) solvers, the outer fixed-point should be noise-aware (e.g., batched/averaged updates, uncertainty-weighted Anderson mixing) with stopping criteria that reflect both residuals and estimated solver variance.}

Within this pipeline, classical pre-processing selects orbitals, constructs effective Hamiltonians, and prepares states; the {quantum} subroutine (variational subspace methods or resource-aware QPE variants) targets the active-space ground or excited states; and post-processing refines and validates results using high-level classical solvers such as semistochastic heat-bath configuration interaction (SHCI) or CCSD(T), including perturbative corrections that mirror the role of triples beyond VQE~\cite{SharmaSHCI2017,li2018fast,li2020accurate,BartlettMusial2007,Windom2024T,Windom2024S}; 
and runtime task delegation across heterogeneous resources (QPU, GPU, CPU) benefits from shared-memory and low-latency synchronization to keep the hybrid loop tight~\cite{BECK202411};

\textcolor{black}{We treat {measurement cost} as a first-class resource: workflows should integrate measurement-reduction pipelines and report total shot budgets per observable (or per desired error bar) alongside logical-gate/$T$ counts~\cite{HuangKernel2021}.}

\textcolor{black}{At the execution level, AI assists this loop end-to-end, including guiding active-space scoring and circuit compilation, accelerating decoder inference and feedback control for QEC, and reducing measurement cost via learning-based estimators (e.g., classical shadows), so that selection, execution, and readout are co-optimized rather than siloed~\cite{Li2021,Duan2022,Alexeev2024AIforQC,HuangKernel2021}.}

Classical assistance can also optimize the quantum subroutine itself. Stabilizer-based bootstrapping leverages classically simulable Clifford circuits to pre-tune parameters and verify compilation/execution paths before running classically intractable circuits~\cite{LiStabilizerBootstrap2024,ravi2022cafqa}, and emerging protocols aim to carry fault-tolerant design ideas into NISQ-era devices for robustness and measurability~\cite{dangwal2025variational}.
\textcolor{black}{As emphasized in Sec.~\ref{sec:validation}, these stabilizer-based checkpoints provide classically verifiable validation of compilation and execution paths before deploying classically intractable circuits. Together, these enable closed-loop control of cost and accuracy.}

\textcolor{black}{Generative circuit synthesis (e.g., GQE) and quantum autoencoders learn low-depth ansätze and compress states, while decoder/scheduling models reduce QEC and latency overheads; on the readout side, learning-based estimators (e.g., classical shadows) cut measurement burden and enable near real-time observable estimation~\cite{nakaji2024generativequantumeigensolvergqe,Alexeev2024AIforQC,HuangKernel2021}.}

\textcolor{black}{Chemical insight (e.g., fragment partitioning) should inform design decisions throughout the stack, from layout-aware transpilation to resource models, guided by community benchmarks and shared roadmaps~\cite{NAP26850}. Community infrastructure is essential for reproducibility and collaboration: use chemistry-aware intermediate representations and automated Hamiltonian-to-circuit translation for performance portability~\cite{claudino2022xacc}, incorporate portable abstractions and standardized modular components, pair these with provenance/FAIR data schemas~\cite{WilkinsonFAIR2016} that link circuits, Hamiltonians, calibration snapshots, decoder outputs, and measurement records to reported observables; maintain public benchmark repositories with well-specified instances and reference outputs~\cite{GSEEBenchmarkProject,Yu2023,Khan2024}, provide cloud/HPC simulators and emulators for large parameter sweeps and AI-in-the-loop studies~\cite{BECK202411}, and distribute workflows as containerized, CI-tested environments to keep results executable over time~\cite{claudino2022xacc}. To measure progress consistently, we anchor development to the community benchmarks and prioritized problem classes outlined in the next section. Ultimately, realizing quantum utility depends on full-stack co-design tightly coupling chemistry, compilers, runtimes, and hardware~\cite{vandam2024endtoendquantumsimulationchemical}.}

\textcolor{black}{\textbf{Hardware Experiment Registry.} To support reproducible benchmarking and AI training across platforms, we advocate a shared registry of real-hardware experiments with a minimal FAIR schema: (i) identifiers and compiled circuits/Hamiltonians (pre/post-transpilation) with versioned seeds and settings~\cite{OpenQASM3,QIRSpec}; (ii) backend metadata and calibration snapshots (e.g., $T_1/T_2$, gate/readout infidelities, crosstalk, code distance $d$ and logical error/cycle when applicable); (iii) runtime logs (session IDs, transpiler passes, scheduling/latency traces) and decoder outputs; (iv) raw/aggregated {measurement records} and shot counts, together with any measurement-reduction used; and (v) links to the benchmark definition, resource ``cost-vector'' (Sec.~\ref{sec:benchmark}), and validation protocol (Sec.~\ref{sec:validation}). Artifacts should be packaged for re-execution (inputs, circuits/Hamiltonians, telemetry, decoder outputs, post-processing scripts) and released under a provenance-preserving format~\cite{W3C-PROV} to enable audit and AI-in-the-loop studies.}

\begin{figure*}
    \centering
    \includegraphics[width=\linewidth]{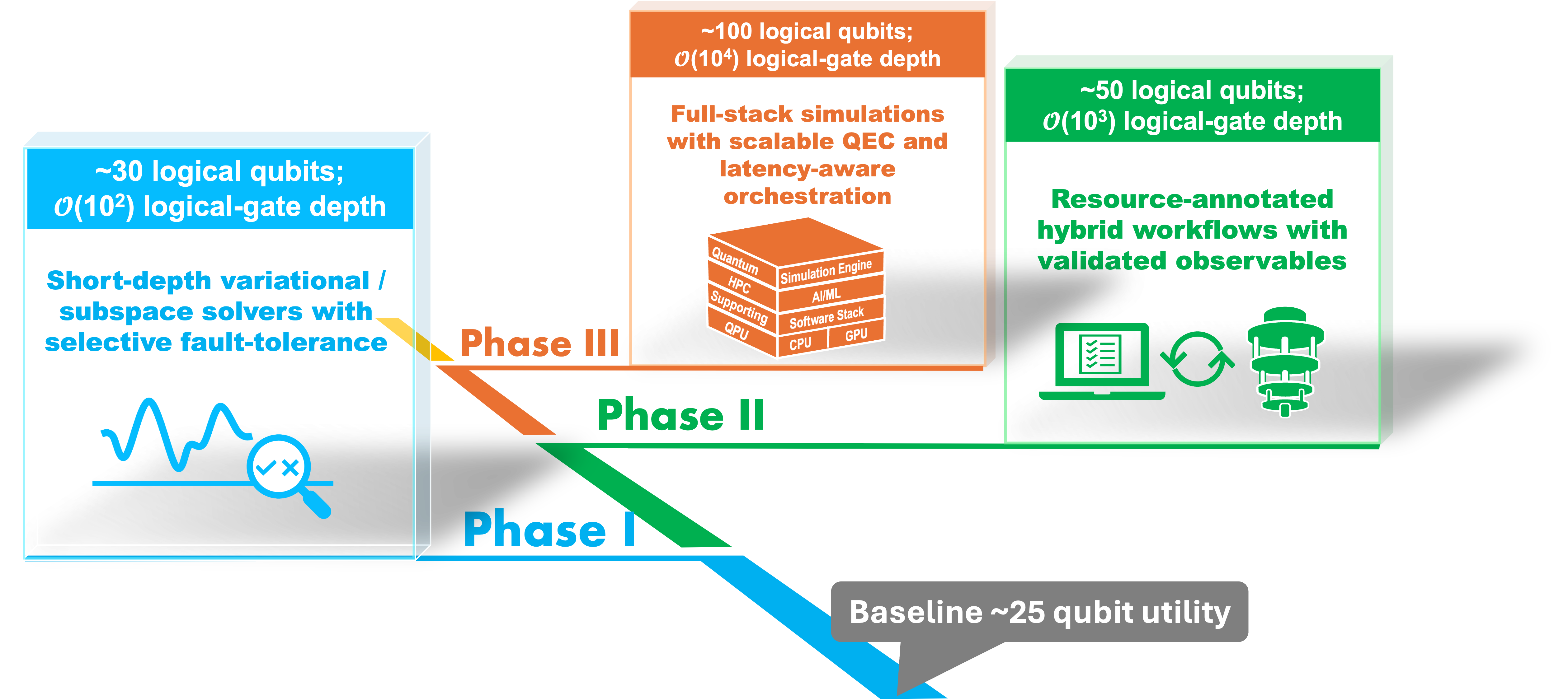}
    \caption{Capability-phased roadmap keyed to performance-based milestones (not calendar dates). Each phase is summarized by indicative logical-qubit counts and logical-gate depth and is evaluated with three metric bundles: QEC/hardware (code distance, logical error rate per cycle), compiler/runtime ($T$-factory/decoder throughput, end-to-end latency), and validation (resource-annotated benchmarks, stated error bars, stabilizer checkpoints). Benchmarks are defined in Sec.~\ref{sec:benchmark}; validation workflow is discussed in Sec.~\ref{sec:roadmap}}
    \label{fig:roadmap}
\end{figure*}

%----------Sec.~V.C

\subsection{\textcolor{black}{Benchmark Design and Problem Prioritization}}\label{sec:class}

\textcolor{black}{\textit{Scope:} Strategies for defining chemically grounded benchmarks and prioritizing problem classes, with explicit resource annotations and environment-aware protocols to guide development.}

\textcolor{black}{The need for chemically grounded benchmark problems will go beyond toy or highly simplified test cases (e.g., minimal systems mainly used for proof-of-concept studies) and instead emphasize chemically realistic, experimentally relevant systems~\cite{McCaskey2019,GSEEBenchmarkProject}.}
Prioritized use cases include challenging electronic-structure problems tied to energy and materials science, as summarized in Table~\ref{tab:combined_benchmarks}.

\textcolor{black}{In addition to chemically motivated benchmarks, stabilizer circuits should be systematically incorporated into benchmark suites as classically verifiable checkpoints for validating compilation/execution paths and resource estimates (see Sec.~\ref{sec:validation}).}
Building on the benchmark framework and Class 0/1/2 definitions introduced in Sec.~\ref{sec:validation}, here we emphasize their role in guiding resource annotation and prioritization. 
\textcolor{black}{Infrastructure benchmarks (Class 0) and challenge benchmarks (Class 1/2) serve complementary purposes within the same program: the former establish reliable execution and comparable metrics, while the latter probe computational difficulty and chemistry-relevant observables beyond current classical limits.}
\textcolor{black}{For Class 1/2 challenge benchmarks, the protocol additionally requires classical-only convergence baselines for the classical components of the workflow (e.g., active-space size, embedding region/bath size, downfolding truncations and tensor-factorization ranks, and environment ensemble/sensitivity), together with the final quantum cost-vector (logical-qubit count, logical depth/$T$-count, post-downfolding term counts/$L_1$ norms, and shot budgets)~\cite{GSEEBenchmarkProject,Alexeev2024AIforQC,McCaskey2019}. These baselines justify the specific instances executed on quantum hardware and enable apples-to-apples comparisons across stacks.}

Notably, the evaluation metrics \textcolor{black}{are} encouraged to include quantitative agreement with experiment, convergence behavior, scaling trajectories, and chemical observables such as excitation energies, spin\textcolor{black}{-}state gaps, and phase boundaries.
\textcolor{black}{Beyond these physics-based metrics, measurement and trainability should be reported explicitly on total shot budgets per observable (or per desired error bar), any measurement-reduction used (e.g., grouping, classical shadows), and trainability diagnostics (gradient norms, optimizer stability/variance scaling) under stated noise and transpilation settings.}
\textcolor{black}{In addition, for embedding and environment-sensitive studies we recommend that benchmarks include: (i) explicit environment metadata (temperature, solvent or dielectric model, ionic composition, cofactors) and an ensemble-averaging protocol; (ii) an uncertainty budget decomposed into ensemble, solver/statistical, and model components; and (iii) emphasis on differential observables (e.g., shifts, gaps, isotope effects) with stated error bars, alongside resource annotations (see Sec.~\ref{sec:class}). Benchmarks should be paired with provenance-rich artifacts (inputs, circuits/Hamiltonians, calibration/decoder outputs) and FAIR metadata for re-execution and auditability (see Sec.~\ref{sec:codesign}).}

%----------Sec.~V.D

\subsection{\textcolor{black}{Strategic Roadmap and Milestones}}\label{sec:roadmap}

\textcolor{black}{\textit{Scope:} A performance-based roadmap linking algorithmic milestones with hardware/QEC, compiler/runtime, and validation metrics, emphasizing reproducibility and community-wide comparability.}

\textcolor{black}{\textbf{What currently limits progress?}
A practical question for the capability-phased roadmap is what most constrains progress over the next few years. In our view, the limiting factors are dual and interdependent. On the {algorithmic} side, key bottlenecks include (i) efficient and portable state preparation for strongly correlated systems and real-time dynamics; (ii) the {measurement overhead} that scales with Hamiltonian variance and target precision (cf. Eq.~(2)), motivating measurement-reduction pipelines (grouping, low-rank factorizations, classical shadows) as first-class resources; (iii) trainability and stability across geometries for variational and subspace methods (mitigated by structured/subspace ansatzes, Krylov/filtered-QPE variants, and error-aware outer loops); and (iv) robust downfolding/embedding with clear commutator truncation and self-consistent outer loops that remain stable when the fragment solver is stochastic, supported by standardized, provenance-rich benchmarks and validation protocols~\cite{Reiher2017,ChildsDynamicsReview2018,Miessen2023QuantumDynamics,Huang2020Shadows,Huang2021shadows,Cerezo2021,VQAChallenges2307,VQAPlateausNPJ2024,VQAClassical2312,KowalskiBaumanDynamics2020,BaumanKowalskiJCP2022,BaumanDownfolding2019,SunDMETReview2016}. On the {hardware} side, realizing sustained computations with $\sim$25–100 logical qubits depends on attaining low logical error rates at practical code distances, adequate decoder/$T$-factory throughput, high-fidelity entangling operations under connectivity constraints, mid-circuit measurement and classical control latencies compatible with hybrid runtimes, and reliable noise models for transpiler and QEC co-design~\cite{IBMRoadmap2022,GoogleSurfaceCode2023,GoogleLogicalProcessor2024,Krinner2022,Bluvstein2024Logical,Fowler2012,RoettelerSvore2018}.}

\textcolor{black}{Crucially, neither front can unlock the roadmap in isolation. Today, the availability and stability of error-corrected logical qubits set the primary gating constraint for end-to-end demonstrations at the 25–100 logical-qubit scale; at the same time, {resource-aware} algorithms (e.g., iterative/statistical/filtered QPE, measurement-efficient subspace methods, and compact effective Hamiltonians via downfolding) directly reduce hardware budgets and accelerate time-to-utility. We therefore anticipate that near-term progress will be paced by {co-design}: aligning algorithmic cost-vectors (logical depth/$T$ count, post-downfolding term counts/$L_1$ norms, shot budgets) with realistic QEC and runtime throughput, integrating HPC/AI for decoding, scheduling, and measurement reduction, and validating against community benchmarks with stated error bars. This interplay, rather than a single bottleneck, sets the slope of progress toward durable quantum utility.}

\textcolor{black}{Rather than date-specific roadmaps, we adopt {performance-based milestones} tied to hardware/QEC, compiler/runtime, and validation metrics (Fig.~\ref{fig:roadmap}, see also Refs.~\citenum{HoeflerAdvantage2023,Huang2021shadows,McCaskey2019,DARPA_QB_Program,PollardBenchQC2025}). Concretely, milestones include: logical error rate per code cycle and attainable code distance (QEC progress), $T$-factory/decoder throughput and end-to-end latency budgets (compiler/runtime), and {validated observable classes} at stated error bars on resource-annotated benchmarks (community readiness, see the previous section for benchmark classes)~\cite{GoogleSurfaceCode2023,Krinner2022}. In Fig.~\ref{fig:roadmap} and Tab.~\ref{tab:phase} we organize the roadmap by capability phases, not dates. Phase I emphasizes short-depth exemplars (e.g., variational/subspace or short-depth QPE) under selective fault-tolerance and reports code distance and logical-error-per-cycle, decoder throughput/latency, and validated observable classes for Class 0 benchmarks. Phases II and III extend these metrics to higher logical-gate depth and qubit counts, with hybrid-workflow benchmarks (Sec.~\ref{sec:benchmark}) and stabilizer-based checkpoints for classically verifiable validation (Sec.~\ref{sec:verify}).}

\textcolor{black}{Reverse-engineering from target applications (e.g., mapping chemical-accuracy thresholds to logical-qubit counts, code distances, and gate fidelities) remains a useful design loop~\cite{vonBurg2021,GSEE2023,watts2024fullerene,otten2024quantum}. Realistic capability claims should report {measurable limits} (trainability, measurement cost, depth at fixed code distance) under stated noise and transpilation settings. For variational circuits, stabilizer-bootstrap and classical surrogates provide classically verifiable checkpoints to quantify trainability and measurement overhead before deployment (see Sec.~\ref{sec:benchmark})~\cite{LiStabilizerBootstrap2024}.}

\textcolor{black}{Progress toward these goals depends on structured collaboration across institutions and disciplines~\cite{NAP26850}.  In practice, this means cultivating shared talent pipelines and organizing multi-institutional consortia aligned with benchmark testbeds (e.g., DOE NQISRC, DARPA Quantum Benchmarking, NSF NQVL/QLCI)~\cite{nqisrc2024,DARPA_QB_Program,nsf_nqvl,nsf_qlci}. Hackathons and challenge problems with specified resource and accuracy targets can accelerate reproducible comparisons and lower entry barriers. Standardized terminology (e.g., logical vs.\ physical qubits, QEC-protected operations) and shared documentation improve cross-domain communication. Equally crucial is the dissemination of {negative results} and {reproducible artifacts} (input decks, circuits, data schemas, calibration snapshots, decoder outputs) via living registries that pair benchmark definitions with reference implementations and provenance, enabling fair cross-platform evaluation as hardware, software, and algorithms co-evolve.}\\

\textcolor{black}{These collaborative structures provide the foundation on which the field can now chart its broader trajectory. We turn to this outlook in Sec.~\ref{sec:conclusion}.}

\begin{table*}[t]
\footnotesize
\setlength{\tabcolsep}{5pt}
\renewcommand{\arraystretch}{1.2}
\caption{\textcolor{black}{Capability-phased milestones (illustrative scales). Phases are defined in Fig.~\ref{fig:roadmap} and keyed to measurable bundles: QEC/hardware, compiler/runtime, and validation. Benchmarks are defined in Sec.~\ref{sec:benchmark}; stabilizer-based validation is discussed in Sec.~\ref{sec:verify}.}}
\label{tab:phase-metrics}
\begin{tabular}{p{2.1cm} p{4.0cm} p{4.3cm} p{4.6cm}}
\hline
\textbf{Phase} & \textbf{QEC / hardware metrics} & \textbf{Compiler / runtime metrics} & \textbf{Validation metrics} \\
\hline
&&&\\
\textbf{I} &
Demonstrate target {code distance} (e.g., $d=5$–$7$) with {logical-error-rate-per-cycle} below a stated threshold. &
Report {end-to-end latency} on critical paths and {decoder throughput}. &
{Validated observable classes} for \textbf{Class 0} entries (e.g., small QM9 subset, alkali hydrides, benzene) with stated error bars; include {stabilizer checkpoints}. \\
&&&\\
\textbf{II} &
Advance to next code-distance point with lower logical error per cycle. &
Provide {compiler/transpiler cost model} and {decoder and $T$-factory throughput} at scale. &
{Hybrid-workflow benchmarks} (e.g., Fe$_2$S$_2$, OLED fragments, model dynamics) with resource annotations; cross-validated vs. high-level classical baselines. \\
&&&\\
\textbf{III} &
Sustained operation at higher code distance with a stable logical-error budget. &
Integrated {GPU–QPU orchestration} targets (latency/throughput); {measurement-reduction pipeline}. &
{Chemically validated observables} for \textbf{Class 1} targets and/or dynamics with stated uncertainty. \\
&&&\\
\hline
\end{tabular}\label{tab:phase}
\end{table*}

%%%%%%%%%%%%%%%%%%Sec.~VI

\section{Conclusion and Outlook}\label{sec:conclusion}

\textcolor{black}{In this concluding section, we synthesize the main insights from the preceding discussions and place them in a broader perspective. The 25--100 logical-qubit regime emerges not only as a technical milestone but also as a proving ground for collaborative strategies that will shape the trajectory of quantum chemistry in the fault-tolerant era.}

%----------Sec.~VI.A

\subsection{A Pragmatic Shift Towards Quantum Utility in Chemistry}

The field of quantum computing for quantum chemistry is undergoing a significant strategic shift. Rather than awaiting the arrival of large-scale, fault-tolerant machines, the community is coalescing around a more pragmatic approach focused on the emerging regime of 25--100 logical qubits~\cite{NAP26850}. This scale is viewed not as a temporary constraint but as a crucial proving ground for demonstrating tangible quantum utility in the coming years. The emphasis has moved towards hybrid, modular, and application-aware strategies that maximize the utility of currently available, limited quantum resources. These strategies often involve techniques like embedding, downfolding, adaptive algorithm design, and importantly, principled integration with classical HPC and AI tools~\cite{Alexeev2024AIforQC, BECK202411, xue2025efficientalgorithmsquantumchemistry}.

Building on this foundation, the central goal has been refined towards achieving ``Quantum Utility,'' demonstrating that a quantum device can provide reliable advantage over the best classical methods on well-chosen, domain-relevant tasks, judged by speed, accuracy, efficiency, and resource costs~\cite{tu2025identifyingpossiblefaulttolerantadvantage}. Rather than general-purpose superiority, this entails prioritizing niche but meaningful problems such as strongly correlated systems, quantum dynamics, and catalysis. Recent demonstrations, like simulating $H_2$ using logical qubits with error detection~\cite{QuantinuumH2Sim2023}, signal tangible progress towards fault tolerance, with quantum chemistry continuing to be a cornerstone application~\cite{Cao2019, McArdle2020, Bauer2020}. 
\textcolor{black}{Equally important, however, is that such demonstrations yield new \emph{scientific insights} (e.g., mechanistic understanding, predictive trends, and emergent quantum behaviors) that remain elusive for classical solvers. Such insights may come, for example, from resolving the role of conical intersections in photochemistry or tracing coherence-driven pathways in enzymatic reactions, providing value beyond numerical accuracy alone and underscoring the broader impact of quantum utility on discovery.}

%----------Sec.~VI.B

\subsection{Co-Design, Collaboration, and Building the Foundation}

Looking ahead, progress hinges on {co-design}: tight coupling between algorithm development and hardware capabilities, coordination across the stages of hybrid simulation workflows, and sustained collaboration among chemistry, computer science, and physics communities~\cite{vandam2024endtoendquantumsimulationchemical,NAP26850,Scholes2025QIS}. Continued innovation across theory, hardware, and software remains essential, with particular emphasis on downfolding and renormalized embeddings, low-depth subspace methods, and emerging fault-tolerant frameworks (e.g., combining fermion and qubit codes)~\cite{KaplanFermionQubit2024}. Focus should also include integrating quantum workflows with efficient classical pre-/post-processing and optimization strategies, leveraging structural diagnostics and fault-tolerant concepts (e.g., stabilizer bootstrapping) even when operating in near-term regimes~\cite{LiStabilizerBootstrap2024}.

In the near term, we prioritize actions that are concrete and measurable: design {chemically relevant, resource-annotated benchmarks} that guide development and quantify progress~\cite{GSEEBenchmarkProject,McCaskey2019}; develop {efficient classical preprocessing and optimization} tailored to quantum subroutines~\cite{LiStabilizerBootstrap2024}; construct {modular, testable hybrid workflows} co-designed with hardware constraints and portable across stacks~\cite{vandam2024endtoendquantumsimulationchemical,claudino2022xacc}; and integrate {AI and HPC} as scalable partners for scheduling, decoding, and measurement reduction~\cite{Alexeev2024AIforQC,BECK202411}. These efforts should be supported by open infrastructure and community benchmarks to ensure that results are reproducible, comparable, and auditable over time.

Ultimately, the work ahead is inherently collaborative. Achieving production-quality quantum chemical simulations will require a shared vision, coordinated investment, and a commitment to open science~\cite{NAP26850,Scholes2025QIS}. The present momentum suggests that practical quantum chemistry is within reach; by grounding progress in co-design, validated benchmarks, and targeted applications, we can build the foundation for durable utility in the 25–100 logical-qubit era and beyond.
\textcolor{black}{In closing, the 25--100 logical-qubit regime should be viewed as the community’s first practical window for demonstrating quantum utility in chemistry, where resource-aware algorithms, auditable benchmarks, and co-designed workflows converge to establish credible, reproducible progress beyond classical limits.}

%%%%%%%%%%%%%%%%%%

\section{Acknowledgements}

This article grew out of discussions at the Workshop on Quantum Computing Applications in Quantum Chemistry, held in April 2025 in Seattle, Washington, USA. The authors gratefully acknowledge the agencies and programs whose support of their individual research efforts made this work possible. In particular, we acknowledge Nathan Baker, Brian Bilodeau, Tamas Gorbe, Hongbin Liu, and David Williams-Young from Microsoft Azure Quantum for their inputs that form fruitful discussions during the workshop. We hope this paper offers a perspective that contributes to ongoing, pressing discussions on the near-future integration of quantum computing and computational chemistry to enable new scientific advances. We thank the Quantum Algorithms and Architecture for Domain Science Initiative (QuAADS), a Laboratory Directed Research and Development (LDRD) program at PNNL, for providing the support for organizing the workshop, and in particular we are grateful to Alison Erickson for her invaluable assistance with workshop logistics and transcription.
A.L., C.L., M.Z., D.C. are supported by the "Embedding QC into Many-body Frameworks for Strongly Correlated Molecular and Materials Systems" project, which is funded by the U.S. Department of Energy, Office of Science, Office of Basic Energy Sciences (BES), the Division of Chemical Sciences, Geosciences, and Biosciences (under award 72689). 
J.L. is supported in part by the University of Pittsburgh, School of Computing and Information, Department of Computer Science, Pitt Cyber, PQI Community Collaboration Awards, John C. Mascaro Faculty Scholar in Sustainability, NASA under award number 80NSSC25M7057, and Fluor Marine Propulsion LLC (U.S. Naval Nuclear Laboratory) under award number 140449-R08. 
B.P. acknowledges the support from the Early Career Research Program by the U.S. Department of Energy, Office of Science, under Grant No. FWP 83466.
MvS is supported by the Computational Chemical Sciences program within the Office of Basic Energy Sciences, U.S. Department of Energy (DOE) under Contract No. DE-AC36-08GO28308.
M.R. is grateful to the Novo Nordisk Foundation for financial support through the Quantum for Life center in Copenhagen/Zurich, NNF20OC0059939. 
Y.Z. acknowledges the support from the Laboratory Directed Research and Development (LDRD) program of Los Alamos National Laboratory (LANL). LANL is operated by Triad National Security, LLC, for the National Nuclear Security Administration of US Department of Energy (contract no. 89233218CNA000001). 

\appendix

\section{Extended details on qumodes and cQED gate implementations}\label{app:qumode}

Qumodes or quantum harmonic oscillators are known for their infinite-dimensional Hilbert space and offer a compelling resource beyond the two-level systems of conventional qubits~\cite{Weedbrook2012}. In practice, the Fock basis of a qumode within an energy cutoff represents a discrete multi-dimensional generalization of a qubit, also known as a qudit~\cite{Wang2020Qudits}. 

Significant progress has been made on quantum devices consisting of qumodes coupled to a qubit, based on the cQED approach~\cite{Blais2021}, where qumodes are realized as microwave cavities dispersively coupled to a superconducting transmon acting as the qubit~\cite{Copetudo2024}. In addition to providing more resources than qubit-only architectures, cQED devices support a diverse range of universal gate sets, including echoed conditional displacement (ECD) with qubit rotations~\cite{Eickbusch2022}, selective number-dependent arbitrary phase (SNAP) gates with displacements and beamsplitters~\cite{Krastanov2015}, and conditional-not displacement gates~\cite{Diringer2024cnot}, among others. 

These gates have recently been explored in applications spanning chemistry~\cite{Dutta2024perspective,Dutta2024EST,Vu2025computational,Dutta2025excitedqumode}, optimization~\cite{Dutta2025solving}, and quantum machine learning~\cite{Smaldone2024quantum}. Hybrid qubit–qumode gates are also difficult to mimic using shallow qubit-only circuits~\cite{Wang2020vibronic,Liu2024review}, and thus offer potential advantages for problems with strong bosonic character (e.g., vibronic simulations~\cite{Wang2020vibronic}) or optimization landscapes favorable to these circuits~\cite{Zhang2024energy}. 

This extended discussion provides the technical backdrop for the concise treatment of qumodes in Sec.~\ref{sec:hardware} of the main text.

\bibliographystyle{unsrtnat}
\bibliography{ref}
\end{document}